\RequirePackage{snapshot}
\documentclass[useAMS,usenatbib]{mn2e}
\usepackage{graphicx}
\usepackage{color}
\usepackage{aas_macros}
\usepackage{amssymb}
\usepackage[nolist]{acronym}

\newcommand{\eagle}[0]{{\sc eagle}}
\newcommand{\gama}{{\sc gama}}

\newcommand{\ltsima}{\mbox{$\; \buildrel < \over \sim \;$}}
\newcommand{\modela}[0]{N}
\newcommand{\modelb}[0]{GI}
\newcommand{\modelc}[0]{GD}
\newcommand{\modeld}[0]{GD+O}

\acrodef{CMD}{colour magnitude diagram}
\acrodef{LF}{luminosity function}
\acrodef{GAMA}{GAlaxy Mass and Assembly}
\acrodef{UKIDSS}{UK Infra-red Digital Sky Survey}
\acrodef{SDSS}{Sloan Digital Sky Survey}
\acrodef{SED}{Spectral energy distribution}

\acrodef{Ref-100}{Ref-L100N1504}
\acrodef{Ref-25}{Ref-L025N0376}
\acrodef{Recal-25}{Recal-L025N0752}

\title[The colour and luminosity properties of {\sc eagle} galaxies]{Colours and luminosities of $z=0.1$ simulated galaxies in the {\sc eagle} simulations}
\author[Trayford et al.]
{James W. Trayford$^{1}$\thanks{E-mail:j.w.trayford@durham.ac.uk (JWT)}, Tom Theuns$^{1}$, 
Richard G. Bower$^1$, Joop Schaye$^{2}$,
\newauthor
 Michelle Furlong $^{1}$, Matthieu Schaller$^1$, Carlos S. Frenk$^1$, Robert A. Crain$^3$, 
\newauthor
Claudio Dalla Vecchia$^{4, 5}$,  Ian G. McCarthy$^3$\\
$^{1}$ Institute for Computational Cosmology, Durham University, South Road, Durham, DH1 3LE\\
$^2$Leiden Observatory, Leiden University, P.O. Box 9513, 2300 RA Leiden, the Netherlands\\
$^3$Astrophysics Research Institute, Liverpool John Moores University, 146 Brownlow Hill, Liverpool L3 5RF  \\
$^4$Instituto de Astrofs\'ica de Canarias, C/ V\'ia L\'actea s/n,38205 La Laguna, Tenerife, Spain \\
$^5$Departamento de Astrofs\'ica, Universidad de La Laguna, Av. del Astrofs\'icasico Franciso S\'anchez s/n, 38206 La Laguna, Tenerife, Spain \\
}

\begin{document}

\date{Submitted 2015 April 15}

\pagerange{\pageref{firstpage}--\pageref{lastpage}} \pubyear{2014}

\maketitle

\label{firstpage}

\begin{abstract}
We calculate the colours and luminosities of redshift z = 0.1 galaxies
from the {\sc eagle} simulation suite using the {\sc galaxev}
population synthesis models. We take into account obscuration by dust
in birth clouds and diffuse ISM using a two-component screen model,
following the prescription of Charlot and Fall. We compare models in
which the dust optical depth is constant to models where it depends on
gas metallicity, gas fraction and orientation. The colours of {\sc eagle}
galaxies for the more sophisticated models are in broad agreement
with those of observed galaxies. In particular, {\sc eagle} produces a
red sequence of passive galaxies and a blue cloud of star forming
galaxies, with approximately the correct fraction of galaxies in each
population and with $g-r$ colours within 0.1 magnitudes of those observed.
Luminosity functions from UV to NIR wavelengths differ from
observations at a level comparable to systematic shifts resulting from
a choice between Petrosian and Kron photometric apertures. Despite the
generally good agreement there are clear discrepancies with
observations. The blue cloud of {\sc eagle} galaxies extends to somewhat higher
luminosities than in the data, consistent with the modest
underestimate of the passive fraction in massive {\sc eagle}
galaxies. There is also a moderate excess of bright blue galaxies
compared to observations. The overall level of agreement with the observed
colour distribution suggests that {\sc eagle} galaxies at z = 0.1 have ages,
metallicities and levels of obscuration that are comparable to those
of observed galaxies.
\end{abstract}


\section{Introduction}
The basic scenario for how galaxies form and evolve is well established: gas accretes onto deepening dark matter potential wells, cools and makes stars. Although this basic paradigm has been accepted for many years \citep[e.g.][]{Rees77, White78, White91}, many important aspects are still poorly understood. For example, the shape of the galaxy stellar mass function (GSMF) and of the halo dark matter mass function are quite different, so a 
simple formation model in which a halo of given mass contains a galaxy whose stellar mass is equal to a fixed fraction of its halo mass is ruled out \citep[e.g.][]{White91, Benson03}. In addition, galaxy surveys have revealed the presence of many correlations between the stellar properties of galaxies and galaxy mass, for example bimodality in the colour-magnitude diagram in the form of a blue cloud of star forming galaxies at lower mass, and a red sequence of mostly passive galaxies at higher mass \citep[{e.g.}][]{baldry04}. Clearly, the dynamics and interaction physics of gas play important roles in determining the properties of galaxies.

It is thought that at the faint end of the GSMF super novae (SNe) quench star formation in small galaxies \citep{Larson74, Dekel86} with reionisation effectively preventing galaxies from forming in dark matter halos below a minimum mass \citep[e.g.][]{Rees86, Efstathiou92, Thoul95, Okamoto08}. Together these processes shape the GSMF at low stellar masses, with dark matter halos containing increasingly feeble galaxies with decreasing halo mass, until most halos remain dark 
\citep{Sawala14}. In contrast, at higher galaxy masses it is thought to be the feedback from accreting black holes (BHs) that introduces a near exponential cut-off in the GSMF \citep{Bower06, Croton06}.

That feedback from star formation and BHs together shape the galaxy stellar mass function is energetically plausible, and semi-analytical models \citep[e.g.][]{Cole00,Henriques13,Gonzalez14,Porter14} and numerical simulations \citep[e.g.][]{Oppenheimer10,Puchwein13,Vogelsberger14,Schaye15} that appeal to these processes are able to produce model GSMFs that compare well to observations at redshift $z\sim 0$. However, neither the semi-analytical models nor these simulations come anywhere near resolving the scales at which SNe and BHs inject energy, and so cannot {\em a priori} compute the net efficiency of the resulting feedback. Simulations therefore need to rely on a phenomenological description of these crucial processes occurring on unresolved  (\lq subgrid\rq) scales, using observations to calibrate the parameters that appear in the subgrid modules. It is then important to quantify the uniqueness and degeneracies in such modelling \citep[][]{Schaye10,Crain15}, while at the same time use very high-resolution simulations \citep[e.g.][]{Hopkins11, Creasey13, Creasey15, Martizzi14, Rosdahl15} to try to bridge the gap between numerically unresolved and resolved scales. 

The {\bf E}volution and {\bf A}ssembly of {\bf G}a{\bf L}axies and their {\bf E}nvironment \citep[\eagle{}][]{Schaye15, Crain15} suite of simulations uses the $z\sim 0.1$ GSMF, together with observations of galaxy sizes, to calibrate the subgrid physics modules. 
The GSMF is however not directly observable, but is inferred from luminosity functions after applying corrections for dust obscuration, 
and using simple stellar population (SSP) synthesis models that
involve assumptions about stellar evolution, star formation histories, metallicity
dependence of stellar emission, etc. With simulations such as \eagle{}
we can take the converse approach, attempting to reproduce the
observational relations by inputting physical quantities tracked by the
simulation. This has the advantage of allowing one to use properties modelled  self-consistently such as gas content, metallicity and age to derive
observable quantities, rather than treating them as free parameters in
\lq SED fitting\rq\ \citep[e.g.][]{Walcher11} to estimate physical properties from observations.
The colours of \eagle{} galaxies are also an important test of the realism of the fiducial \eagle{} model.

In this paper we examine to what extent mock luminosities computed from \eagle{} galaxies using SSP models and a simple correction for dust obscuration reproduce the observed luminosity functions (in a range of broad bands), as well as galaxy colours. The aim is twofold: to provide a test of the realism of \eagle{}, but also to test at some level whether the procedure of going from luminosity to stellar mass is reliable, as investigated in a recent paper by \cite{Torrey14} using SED fitting of galaxies from the {\sc Illustris} simulation \citep{Vogelsberger14}. It is perfectly possible that \eagle{} galaxies have the wrong colours but the right stellar masses and stellar ages if errors in dust modelling are severe. However, if masses {\em and} colours agree with the data, then this increases our confidence that we can use \eagle{} to investigate, for example, the origin of the observed bimodality of galaxy colours, or the nature of the galaxies that lie in between the red sequence and blue cloud in a colour-magnitude diagram. 

In what follows we compare to photometric data from the {\it Galaxy And Mass Assembly} survey \citep[\gama{}][]{Driver11}. This is a
spectroscopic follow-up based on photometric data from the {\it Sloan Digital Sky Survey}
(SDSS) and the {\it UK Infrared Digital Sky Survey} (UKIDSS), with
details on the targeting and star-galaxy separation in \citet{Baldry10} and on the GAMA-processed photometry, including matched aperture photometry from
$u$ to $K$ in \citet{Hill11}. The \gama{} survey has been designed for
high uniform spectroscopic completeness \citep{Robotham10}
and provides accurate redshifts \citep[using AUTOz;][]{Baldry14} for a
catalogue of $\sim 190,000$ galaxies, as presented by \citet{Taylor14}.

This paper is organised as follows. We begin with an overview of the \eagle{} simulations, with emphasis on those aspects that are most relevant
for the SSP modelling. In \S \ref{sec:phot} we detail the development of our photometric model, concentrating on emission and absorption in \S \ref{sec:emission} \& \S  \ref{sec:absorption} respectively. This model is applied to yield an optical colour-magnitude diagram (CMD) and multi-band luminosity functions (LFs) for galaxies, which
are plotted and discussed in \S \ref{sec:results}. We discuss our findings in \S \ref{sec:disc} and summarise in \S \ref{sec:conclusion}.

\section[]{The {\bf\sc EAGLE} simulations}
\label{sec:sim}
\begin{table} 
\begin{center}
\caption{Numerical parameters of those simulations of the \eagle{} suite that are used in this paper. From
left to right: simulation identifier, side length of cubic volume $L$ in
co-moving Mpc (cMpc), initial mass $m_{\rm g}$ of baryonic particles, Plummer-equivalent
gravitational softening $\epsilon_{\rm prop}$ at
redshift $z=0$ in proper kpc (pkpc).} 
\label{tab:sims}
\begin{tabular}{lrrr}
\hline
Name & $L$ & $m_{\rm g}$ & $\epsilon_{\rm prop}(z=0)$ \\  
& cMpc & ${\rm M}_\odot$ & pkpc\\
\hline 
\ac{Ref-25} &  25 & $1.81\times 10^6$ & 0.70\\
\ac{Recal-25} &  25 & $2.26\times 10^5$ & 0.35\\
\ac{Ref-100} & 100 & $1.81\times 10^6$ &0.70\\
\hline
\end{tabular}
\end{center}
\end{table}

Full details of the \eagle{} simulations can be found in \cite{Schaye15} and \cite{Crain15} (hereafter S15 and C15 respectively); here we give only a brief overview. The \eagle{} simulation comprises a suite of cosmological hydrodynamical simulations of periodic cubic volumes performed with the {\sc Gadget-3} TreeSPH code (which is an update of the {\sc Gadget-2} code last described by \citealt{Springel05}). Simulations were performed for a range of volumes and numerical resolutions. Here we concentrate on the reference model, using simulations at different resolution to quantify numerical convergence. The reference model assumes a $\Lambda$CDM cosmology with parameters derived from the initial {\it Planck} \citep{Planck} satellite data release ($\Omega_{\rm b} = 0.0482$, $\Omega_{\rm dark} = 0.2588$, $\Omega_\Lambda = 0.693$ and $h = 0.6777$, where $H_0 = 100\; h$ km s$^{-1}$ Mpc$^{-1}$). Relevant properties are listed in Table \ref{tab:sims}.

We modified the treatment of hydrodynamics in {\sc Gadget-3} to use the conservative pressure-entropy SPH formulation of \citet{Hopkins13}, the artificial viscosity switch introduced by \cite{Cullen10}, an artificial conduction switch inspired by \citet{Price08}, the $C^2$ kernel from \citet{Wendland95}  and the timestep limiter of \citet{Durier12}.
Motivation and tests of this \lq {\sc Anarchy}\rq\ version of SPH are presented in Dalla Vecchia (in preparation, see also Appendix A of S15), whereas Schaller et al (in preparation) examine the effects of using this modified version of SPH on galaxy properties.

A crucial aspect of \eagle{} is that the parameters describing the subgrid modules have been calibrated on the observed $z\sim 0$ GSMF and galaxy sizes. This good agreement extends to many other observables that were not considered during the calibration, such as specific star formation rates (S15), the evolution of the GSMF \citep{Furlong14}, molecular hydrogen fractions \citep{Lagos15}, and absorption by intergalactic metals and neutral hydrogen \citep[S15][]{Rahmati15}.

We now briefly describe those aspects of the subgrid model most relevant for this paper.

\subsection{{\sc\bf EAGLE} Subgrid physics}
\label{sec:bary}
The subgrid modules of \eagle{} are partly inspired by the {\sc owls} and {\sc gimic} simulations \citep{Schaye10,
  Crain09}. Star formation is implemented as described in \citet{Schaye08}: above the metallicity-dependent star formation threshold of \cite{Schaye04}, cold enough particles ($T \sim 10^4$K) are converted to star particles stochastically at a pressure-dependent rate that reproduces the observed Kennicutt-Schmidt star formation law. Each simulation star particle is assumed to represent a coeval population of stars formed with a \cite{Chabrier03} IMF, comprising stars with masses in the range $[0.1,100]~{\rm M}_\odot$. 

Stellar evolution is implemented as described in S15 and \citet{Wiersma09a}. We follow the production and release into the interstellar medium of 11 elements from three channels of stellar evolution (AGB stars, type I and type II SNe) using metallicity-dependent stellar lifetimes and stellar yields. We also
track a separate \lq total\rq\ metallicity (the mass fraction of
elements more massive than helium), $Z$, to account for elements not tracked explicitly. When a gas particle is converted into a star particle it inherits the gas particle's abundances. In addition to a particle metallicity, the simulation tracks smoothed metallicities which are computed using the SPH formalism to partly remedy the absence of mixing in the calculation (see \citealt{Wiersma09a} for motivation). The ages, masses and metallicities of the star particles are the main ingredients of the simple stellar population (SSPs) models used below.

Radiative cooling and photo-heating in the presence of an optically thin UV/X-ray background, as computed by  \citet{Haardt01}, is accounted for as described in \citet{Wiersma09b}. The crucial processes of feedback from star formation is implemented by stochastically heating particles by a fixed temperature increment, as described and motivated in \citet{DallaVecchia12}, and adapted for the simulation as described in S15 and C15. The formation of supermassive black holes and gas accretion onto them is implemented as in \cite{Booth09}, with modifications described in \citet{RosasGuevara13} and S15, using a single feedback mode.

\subsection{Identifying galaxies}
\label{sect:galaxies}
To group star particles into \lq galaxies\rq, we proceed as follows. We begin by identifying dark matter halos using the
friends-of-friends algorithm \citep{Davis85} with a linking length of
0.2 times the mean dark matter inter-particle spacing to identify
regions that are overdense by a factor of $\sim 200$
\citep{Lacey94}. We then use the {\sc subfind} algorithm
\citep{Springel01,Dolag09} to identify self-bound substructures (subhalos) within
halos of dark matter, stars and gas, which we identify with
galaxies. Massive galaxies in \eagle{} have extended density
profiles. To assign luminosities to mock galaxies, we only include
light emitted within a sphere of radius 30~ proper kpc (pkpc), centred on the minimum of potential of each subhalo. The motivation for the choice of aperture is discussed further in section~\ref{sec:apt} and also in S15.

\section{Photometry}
\label{sec:phot}
This section explains how we compute luminosities and colours for the simulated \eagle{} galaxies.
We begin by modelling luminosities of star particles, taking into account their ages and metallicities using population synthesis (section ~\ref{sec:emission}), the photometric system used to calculate magnitudes for direct comparison to observation (section~\ref{sec:ugriz}), and the effects of dust absorption (section~\ref{sec:absorption}). The results of this section are summarised in Fig.~\ref{fig:cmddev}, in which we plot histograms of $g-r$ colours of \eagle{} galaxies in narrow stellar mass bins for different models, ranging from a simple emission model without dust, to a model including a multivariate treatment for dust.

\subsection{Source Modelling}
\label{sec:emission}
Below we compute luminosities of \eagle{} galaxies from the ultraviolet (UV) to the near infrared (NIR). We limit ourselves to modelling stars, neglecting both nebular emission lines and light from AGN. Light absorbed by dust is assumed to be re-emitted in the far infrared which we do not study in this paper. As we also neglect scattering by dust, we treat individual wavelength bands independently. This approximation is supported by observations showing that scattering is a negligible contributor to the observed attenuation curve in galaxies \citep[eg.][]{Fischera03}.

Population synthesis models provide spectra for a discrete range of {\it simple stellar populations}
(SSPs) \citep[e.g.][]{bc03, m05, sb99}. These SSPs represent populations of stars characterised 
by their total initial mass, formation time, and composition while assuming some stellar IMF. By decomposing the stellar component
of a galaxy into a superposition of  SSPs, the spectral energy distribution of an entire galaxy can be approximated.
We treat each \eagle{} star particle as an SSP with given initial stellar mass, metallicity and age.

\subsubsection{SSP Ingredients}
\label{sec:ingredients}
Given our implementation of star formation, where gas particles are wholly converted into star particles, the typical mass of an \eagle{} star particle is $\sim 10^6{\rm M}_\odot$. Stars are assumed to form with a \cite{Chabrier03} IMF (for consistency with the evolutionary models used in \eagle{}), and they inherit the SPH-smoothed metallicity, $Z$, from their parent gas particle. The mass of a newly formed star particle is purely set by numerical resolution; the particle should not be thought of as representing a star cluster. In fact, $10^6$ ${\rm M}_\odot$ is higher than the stellar mass formed in giant H{\sc ii} regions \citep[e.g.][]{Relano09, Zaragoza14}. This poor sampling of star formation can adversely affect luminosities of \eagle{} galaxies. Indeed, a single recently formed star particle can affect the colour of a galaxy. We try to mitigate this numerical artifact by employing a finer sampling of the star formation history of recently formed stars, as described in Appendix~\ref{ap:adapt}. We note that this has only a small effect for optical colours and thus for the results presented here.

The metallicity of stars affects their colours resulting in the well-known {\em age-metallicity} degeneracy \citep[e.g.][]{Worthey94}.
In addition, $Z$ affects stellar evolution leading to differences between models, particularly for the AGB phase \citep[e.g][]{Inoue12, Stancliffe07}. In addition, metallicity of stars in \eagle{} galaxies is set by the intricate interaction between enrichment of the ISM, gas accretion, and the extent to which galactic winds transport metals into the galaxy's circum- and intergalactic medium. The details of such metal mixing are still poorly understood and numerically challenging to model. Hence, there is no a priori guarantee that \eagle{} yields realistic stellar metallicities.

The stellar mass - metallicity ($M_\star-Z_\star$) relations provide a useful way of characterising stellar abundances
as a function of galaxy mass, and as shown in S15, the \ac{Ref-100} model yields stellar and gas-phase metallicities consistent with observations \citep{Tremonti04, Gallazzi05, Zahid14} for stellar masses $M_\star\gtrsim 10^{10}{\rm M}_\odot$. However, lower-mass galaxies in \eagle{} tend to be more metal-rich than observed, with numerical resolution playing a role: the higher-resolution \ac{Recal-25} simulation agrees with the data for $M_\star \lesssim 10^9 {\rm M}_\odot$. It should also be noted that there are large systematics on the observed mass-metallicity relations \citep[e.g.][]{Kewley08}. We investigate the impact of stellar metallicity ($Z_\star$) on \eagle{} colours in more detail in Appendix~\ref{ap:gallazzi}. In our analysis, we use the \eagle{} SPH-smoothed metallicities \citep{Wiersma09b} for each particle, which yield less noisy estimates of $Z_\star$.

\subsubsection{Stellar Population Synthesis (SPS) Modelling}
\label{sec:sps}
We adopt the {\sc galaxev} population synthesis models of \citet{bc03}, which provide the spectral energy distribution (SED) per unit initial stellar mass of a SSP for a discrete grid of ages (ranging from $t = 10^5$ to $2 \times 10^{10}$~yr) and metallicities (ranging from $Z_\star=10^{-4}$ to 0.05). We compute the SED for each stellar particle by interpolating the {\sc galaxev} tracks
logarithmically in age and $Z_\star$, and multiplying by the initial
stellar mass. Values of $Z_\star$ in the simulation span a wider range
than the models of \citet{bc03} encompass, and we extrapolate the
model to predict colours for such extreme metallicities\footnote{We find that
the effect of extrapolating metallicities, as opposed to clipping metallicities to
that of the highest metallicity {\sc galaxev} spectra, has a
negligible effect on our results.}. We note that the more conservative approach of not extrapolating introduces a bias.

The {\sc galaxev} spectra are based on stellar emission alone. These models are widely used, and have been shown to fit the local galaxy population in the optical bands with reasonable star formation and enrichment histories when used in conjunction with a dust model \citep[e.g.][]{CF00, Cole00}. The choice of population synthesis model has been shown to be
largely unimportant for low-redshift galaxy populations, especially in optical bands \citep[e.g.][]{GonzalezPerez14}. The effect of different models
\citep[e.g.][]{m05, Conroy09}, and in particular the more uncertain impact of thermally-pulsing AGB stars (TP-AGB) on the SED, should however be considered when surveying higher redshift ($z\gtrsim 1$) galaxy populations \citep[e.g.][]{m05, Gonzalez14}. 

The \citet{bc03} models specify $Z_\star$ values as absolute metal-mass fractions, where $Z_\star$ affects the colours of stars through its impact on stellar structure and evolution - for example via the opacity, equation of state and mean molecular weight - and on stellar atmosphere models. Even so, the metallicity of the Sun ($Z_\odot$) {\em does} enter the population synthesis models because some parameters, such as the mixing length, are calibrated based on solar observables \citep{Bressan93}. 
Recent literature determinations of $Z_\odot$ have shown significant variability, with a minimum of $Z_\odot \sim 0.0126$ \citep[e.g.][]{Asplund04} from the traditional value of $Z_\odot=0.02$ assumed in \citet{bc03}. Although the \eagle{} simulations do not make use of any particular solar abundance pattern or $Z_\odot$ value, a relatively low value of $Z_\odot=0.0127$ \citep{Allende01} has been assumed in analysis for consistency with \citet{Wiersma09b} (S15). The variation of $Z_\odot$ generally results from a different determination of the abundance of some important element, such as O, N, C or Fe which also implies a variation of the abundance partition in the Solar model. In principle one should use {\sc galaxev} SSP models with an abundance partition consistent with the assumed value of $Z_\odot$, and take into account effects arising from the different mixing length calibration, to compute colours self-consistently. For now we neglect any such changes and use the original {\sc galaxev} SEDs, as the effects of this change on broad-band colours are expected to be small, as long as one makes use of the absolute value of the metallicity (Bressan 2014 \textit{private comm.})\footnote{We are grateful to S. Charlot and A. Bressan for their detailed explanation of the impact of $Z_\odot$ on the {\sc galaxev} model.}.

\subsubsection{Photometric System}
\label{sec:ugriz}

Given the SED for each star particle in the simulation, and a model for attenuation due to dust as a function of wavelength, we could compute the SED for each galaxy, and calculate a broad-band magnitude by convolving with a broad-band filter. Here we use a much simpler method, namely we {\em first} convolve the {\sc galaxev} spectra (for each value of age and metallicity) with broad-band filters to obtain \lq un-obscured\rq\ broad-band luminosities. We use these to obtain a broad-band luminosity for an \eagle{} galaxy. We only {\em then} take into account dust attenuation (as described below). If, as we assume, the wavelength dependence of the dust attenuation is not very strong ({\em i.e.} the optical depth does not vary strongly over the wavelength extent of the filter), then these two approaches yield very similar results. For the dust models discussed below, we verified that this is indeed the case (section ~\ref{sec:absorption}).

We use the {\it ugrizYJHK} photometric system for optical and near infrared photometry, to enable a direct comparison 
to the {\sc gama} survey \citep[described in][]{Driver11, Hill11}. This survey is based on the photometry of SDSS \citep[technical description in][]{York00} and  UKIDSS, \citep[technical description in][]{Lawrence07}.
Note that when calculating photometry below, filter transmission curves include atmospheric absorption to enable a like-for-like comparison of simulation and observation. All magnitudes are calculated as rest-frame and absolute in the AB-system \citep{Oke74} in which the apparent magnitude $m_{\rm AB}$ is defined as  
\begin{equation}
m_{\rm AB}\equiv -2.5\log_{10}(F_\nu)-48.6\, ,
\end{equation}
where $F_\nu$ is the isophotal flux density (in cgs units) in a particular band \citep[e.g.][]{Tokunaga05}.

\subsubsection{Choice of aperture}
\label{sec:apt}
Individual \eagle{} \lq galaxies\rq\ are identified as described in \ref{sect:galaxies}. We select galaxies with at least 100 star particles, whose stellar mass is reasonably well converged numerically (S15). The line of sight is chosen consistently to lie along an axis of our simulation coordinates, yielding an essentially randomised orientation for each galaxy. 

Massive galaxies ($M_\star \gtrsim 10^{10.5}{\rm M}_\odot$) in \eagle{} have up to $\sim40\%$ of their stellar mass in an extended halo beyond 30 pkpc of the galaxy centre (\lq intra-cluster light\rq\, since most of these massive galaxies are at the centre of a group or cluster). Observationally, such galaxies also tend to have extended light distributions and, unsurprisingly, the luminosity assigned to them depends on how such light is taken into account \cite[e.g.][]{Bernardi13}. We apply a constant aperture of 30~pkpc centred on the minimum of the gravitational potential of each subhalo for measuring the total luminosity, $L$, of a galaxy. The luminosity and colour of a galaxy with a significant intra-cluster light component do depend on whether we apply a 2D aperture 
or a 3D aperture.
This change in colour is due to colour gradients but also due to the inclusion or exclusion of small blue satellite objects below the significance of those identified by {\sc subfind}. This is a similar issue to that encountered when isolating galaxies in astronomical data using software such as {\sc Sextractor} \citep{Jackson10}. We choose to apply a 3D spherical aperture, consistent with our previous analysis (S15, \citet{Furlong14}). Such an aperture is shown in S15 to yield similar stellar masses to a Petrosian aperture, often used in observational studies. 

The aperture definition is not standardised in observations, and can make a difference when a considerable fraction of stellar material lies outside
the aperture. This is illustrated for the Kron and Petrosian apertures in \citet{Driver12}, where luminous galaxies with high S\'{e}rsic indices yield different magnitudes. Similarly, when analysing our simulations, the luminosities of \eagle{} galaxies with $M_\star \gtrsim 10^{11}{\rm M}_\odot$ are sensitive to the exact choice of aperture size. However, this is not the case for lower mass galaxies, for which the fraction of light in an extended halo is much lower.

\subsubsection{Model \modela{}}
\label{modelE}

The procedure for obtaining \eagle{} galaxy photometry outlined above (sections \ref{sec:ingredients}-\ref{sec:apt}) provides a model with no consideration of dust effects. This is hereafter referred to as model \modela{}. Model \modela{} provides a basis for comparison with photometry corrected for dust attenuation, as described below (sections \ref{sec:static-dust}-\ref{modelE-MCFO}). Panel a) of Fig. \ref{fig:cmddev} shows the $g-r$ colour distribution of \eagle{} galaxies for this model. 

\subsection{Dust model}
\label{sec:absorption}
We develop a simple {\it empirically calibrated} model for dust absorption, as opposed to a more physical modelling using ray-tracing, which we will present elsewhere. One advantage of such a model is that we can easily disentangle the effects of dust from those
of the SPS modelling on galaxy colours.  In addition, if we model dust attenuation using galaxy parameters that
are provided by \eagle{} but can also be inferred through observation, then we may calibrate the reddening of \eagle{} galaxy colours empirically to reproduce observed trends. Keeping our dust model parametrisation independent of certain quantities, such as the gas distribution, also allows us more freedom to investigate the extent to which certain assumptions affect galaxy colours.

In our modelling, dust corrections are applied as a multiplicative factor that reduces a given broad-band luminosity.
This factor is estimated at the effective wavelength of each filter (SDSS filter parameters taken from \citeauthor{SDSSfilters} 2010, UKIRT filter parameters taken from \citeauthor{UKIRTfilters} 2006), for a given dust prescription (neglecting changes in the dust opacity within a broad-band filter is a good approximation provided the dust model has a smooth wavelength dependence). In this way, dust obscuration depends on the subhalo properties of a galaxy alone, and we avoid handling entire SEDs and generating a new interpolation grid for each galaxy. This makes the analysis process very efficient. The approximation that reddening can be applied {\em after} the spectrum has already passed through a filter affects the $g-r$ optical colours by $\lesssim$ 0.02 mag over the whole interpolation grid for the constant optical depth model discussed below.

\subsubsection{Model \modelb{}: Galaxy-independent dust model}
\label{modelE-CF}
\label{sec:static-dust}

We begin by discussing the simplest dust model introduced by  \citet{CF00}, hereafter referred to as CF.
This model includes two contributions to the dust optical depth in a galaxy ($\hat{\tau}_{\rm d}$): ({\em i}) a transient component due to dust in stellar birth clouds ($\hat{\tau}_{\rm bc}$), and ({\em ii}) a constant dust screen that represents dust in the ISM ($\hat{\tau}_{\rm ism}$). The transmission $T$ of this model is

\begin{equation}
T(\lambda, t')=\cases
{\exp\left(-\left[ \hat{\tau}_{\rm bc} + \hat{\tau}_{\rm ism}\right]\left(\frac{\lambda}{\lambda_v}\right)^{-0.7}\right)&, for
$t'\leq t_{\rm disp}\,$,\cr
\exp\left(-\hat{\tau}_{\rm ism} \left(\frac{\lambda}{\lambda_v}\right)^{-0.7}\right)&, for $t'>t_{\rm disp}\,$.\cr}
\label{eq:CF}
\end{equation}

Here, $t'$ is the stellar particle's age, $t_{\rm disp}$ is the dispersal time of the stellar birth cloud, $\lambda$ is the wavelength of light, and $\hat\tau_{\rm ism}$ and $\hat\tau_{\rm bc}$ characterise the total dust optical depth (due to ISM and birth-cloud, respectively), at wavelength $\lambda_v$. When a model SED of a galaxy is fit to an observed galaxy, $\hat{\tau}_{\rm ism}$ and $\hat{\tau}_{\rm bc}$ can be used as fitting parameters to model dust \citep[e.g.][]{daCunha08}. Alternatively, these parameters can be assigned constant values to model dust for a given population of galaxies as in \citet{bc03}, but this does then not allow for variations between galaxies.

As a first approximation we simply take $\hat\tau_{\rm ism}$ and $\hat\tau_{\rm bc}$ to be constants,\begin{eqnarray}
\hat{\tau}_{\rm bc} &=& 0.67\nonumber\\
\hat{\tau}_{\rm ism} &=& 0.33,
\label{eq:tau}
\end{eqnarray}
with $\lambda_v=5500$~\AA\ and $t_{\rm disp}=10^7$~yr, which were calibrated to fit starburst galaxies and were used for the recent analysis of the {\sc illustris} simulations \citep{Vogelsberger14} by \citet{Genel14} and \citet{Torrey14}. With the optical depths fixed, the colours of an \eagle{} galaxy will only depend on the SSP modelling. Such an approximation was also used in the development of the {\sc  galaxev} model, where it was shown to work well when SED fitting a subset of the  SDSS survey at $z=0.1$ \citep{bc03}. The effect of this simple dust model on $g-r$ colours can be seen by comparing panels a and b in Fig.~\ref{fig:cmddev}, and is discussed in more detail below.

In this simple model (model \modelb{}), the strength of the two screen components are fixed for each galaxy (and hence do not depend on e.g. its gas mass or metallicity) and are also independent of orientation. We discuss refinements of the dust model next.

\subsubsection{Model \modelc{}: ISM-dependent dust model}
\label{model:E-MCF}
\label{sec:new-dust}

To account for variations in metal enrichment in the interstellar medium (ISM) of galaxies, we use the mass-weighted SPH-smoothed metallicity \citep{Wiersma09b} calculated for star-forming gas in each \eagle{} subhalo, $Z_{\rm SF}$. This metallicity calculation is chosen to imitate observational measurement techniques \citep{Tremonti04, Zahid14}. As demonstrated in S15, the mass-metallicity relations in \eagle{} are significantly affected by resolution, with the \ac{Recal-25} simulation showing better agreement with the observed $Z_g-M_\star$ relation from \citet{Tremonti04} than \ac{Ref-100} for $M_\star \ll 10^{10} {\rm M}_\odot$\footnote{Note, however, that the observed mass-metallicity relation suffers from calibration uncertainties that exceed the difference between \citet{Tremonti04} and \ac{Ref-100} \citep{Kewley08} and that the more recent re-analysis by \citet{Zahid14} falls in between \ac{Ref-100} and \citet{Tremonti04} (see figure 13 of S15).}. Near the knee of the mass function, however, the \eagle{} mass-metallicity relation agrees well with observation (see S15).

We take the analytic expression for the $M_\star-Z_g$ mass-metallicity relation at $z=0.1$ of \citet{Zahid14}, $Z_{\rm Z14}(M_\star)$, evaluated at the Milky Way stellar mass, 
$M_{\rm MW}=6.43\times10^{10}~{\rm {\rm M}_\odot}$ \citep{McMillan11}, as the ISM metallicity represented by the optical depth values of  Eq.~(\ref{eq:CF}). Assuming optical depth is proportional to metallicity, we then scale the optical depths $\hat\tau_{\rm bc}$ and $\hat\tau_{\rm ISM}$ that appear in Eq.~(\ref{eq:CF}) by the factor
\begin{eqnarray}
\hat\tau_{\rm bc}&\rightarrow  & {Z_{\rm SF}\over Z_{\rm Z14}(M_\star=M_{\rm MW})}\,\hat\tau_{\rm bc}\nonumber\\
\hat\tau_{\rm ism}&\rightarrow &  {Z_{\rm SF}\over Z_{\rm Z14}(M_\star=M_{\rm MW})}\,\hat\tau_{\rm ism},
\label{eq:tauZ}
\end{eqnarray}
for each \eagle{} galaxy.

Making the dust optical depth depend on metallicity is physically well motivated, as it is indicative of the fraction of ISM mass in dust grains. Therefore, we must also take account of the gas mass to quantify how much dust is available for obscuration.
We do so by making the dust optical depth dependent on the {\em cold gas} mass - but still neglect how that gas is distributed.

We approximate the cold gas mass, $M_{\rm ISM}$, by the mass in star-forming gas, which in \eagle{} means gas above a given metallicity-dependent density threshold and below a given temperature (see S15). We then scale the birth cloud and ISM dust optical depths by the ratio of $M_{\rm ISM}$ for the galaxy over the value for the Milky Way \citep[which we take to be $10$~per cent of $M_\star$;][]{Mo10, McMillan11}. This scaling is derived from observations indicating that optical depths approximately scale with the overall gas surface density ($\Sigma_g$) of galaxies \citep[e.g.][]{Grootes13, Boquien13}. By taking $\Sigma_g$ as approximately $ \propto M_{\rm ISM} / {R_\star}^2$, and since the $z=0.1$ stellar mass-size relation is relatively flat for  both \eagle{} (S15) and observed galaxies \citep{Shen03}, we approximate that $\tau \propto \Sigma_g \propto M_{\rm ISM}$. Neglecting the ${R_\star}^{-2}$ dependence maintains a relatively simple parametrisation, and appears to have little effect on galaxy colours, due to the limited mass range over which reddening is significant.

Such a scaling has the desired effect of reddening gas-rich spiral galaxies more than gas-poor elliptical galaxies at the same $M_\star$. The galaxy $g-r$ colour distributions for the model including metallicity and gas fraction dependent reddening (model \modelc{}) are shown in Fig.~\ref{fig:cmddev}c. For comparison, we also show the $g-r$ colour distributions for \eagle{} galaxies where the value of $Z_{\rm Z14}(M_\star)$ is used in Eq.~(\ref{eq:tauZ}), instead of $Z_{\rm SF}$. Because low $M_{\rm ISM}$ values provide low optical depths for the stellar mass range where \eagle{} and observed mass-metallicity relations differ ($M_\star \ll 10^{10} {\rm M}_\odot$), both treatments produce similar distributions.

\subsubsection{Model \modeld{}: ISM-dependent dust model with orientation effects}
\label{modelE-MCFO}

Finally, we represent the dependence of obscuration on orientation with a simple toy model. We assume the dust geometry to be an oblate spheroid, with major to minor axial ratio of $q=a/b=0.2$. This $q$ value is commonly used to represent an axial ratio typical of the intrinsic stellar distribution in disc galaxies \citep[e.g.][]{Tully77}. We assign to each \eagle{} galaxy an orientation $w=\cos(\theta)$, where $\theta$ is the angle between the galaxy's minor axis and the line of sight. To obtain a random orientation we randomly sample $w$ from a uniform distribution over the interval $[-1,1]$. The line-of-sight depth through the dust spheroid is then

\begin{equation}
  \label{eq:ellipse}
    l(w) = a \frac{q}{\sqrt{q^2 + (1 - q^2)w^2}}.
\end{equation} 
We then scale $\hat\tau_{\rm ISM}$ as
\begin{equation}
\hat\tau_{\rm ism} \rightarrow {l(w)\over\langle l\rangle}\,\hat\tau_{\rm ism}\,,
\end{equation}
so that the mean optical depth does not change. This scaling reduces the amount of dust obscuration for most galaxies by a small amount, yet increases $\tau$ by a factor $\sim 2$ for a small number of \lq edge-on\rq\ systems. We assume a value of $q$ that is appropriate for discs, but we note that elliptical galaxies - provided they have little cold gas - are not strongly reddened anyway, hence this orientation correction is not important for them. The $g-r$ colour histograms including orientation effects are shown as model \modeld{} in Figure \ref{fig:cmddev}d. We also show the colour distributions produced using more oblate geometries, with axial ratio values $q = 0.1$ and $q = 0.02$, for comparison.

\section{Results}
\label{sec:results}

In this section we examine the effects of dust modelling on the
colours, luminosities, and colour-magnitude diagrams of \eagle{} galaxies
taken from the \ac{Ref-100} and \ac{Recal-25} models at redshift $z=0.1$. 

\subsection{Galaxy colours as a function of stellar mass}
\begin{figure*}
  \centering
  \begin{minipage}{0.47\textwidth}
  \includegraphics[width=\textwidth]{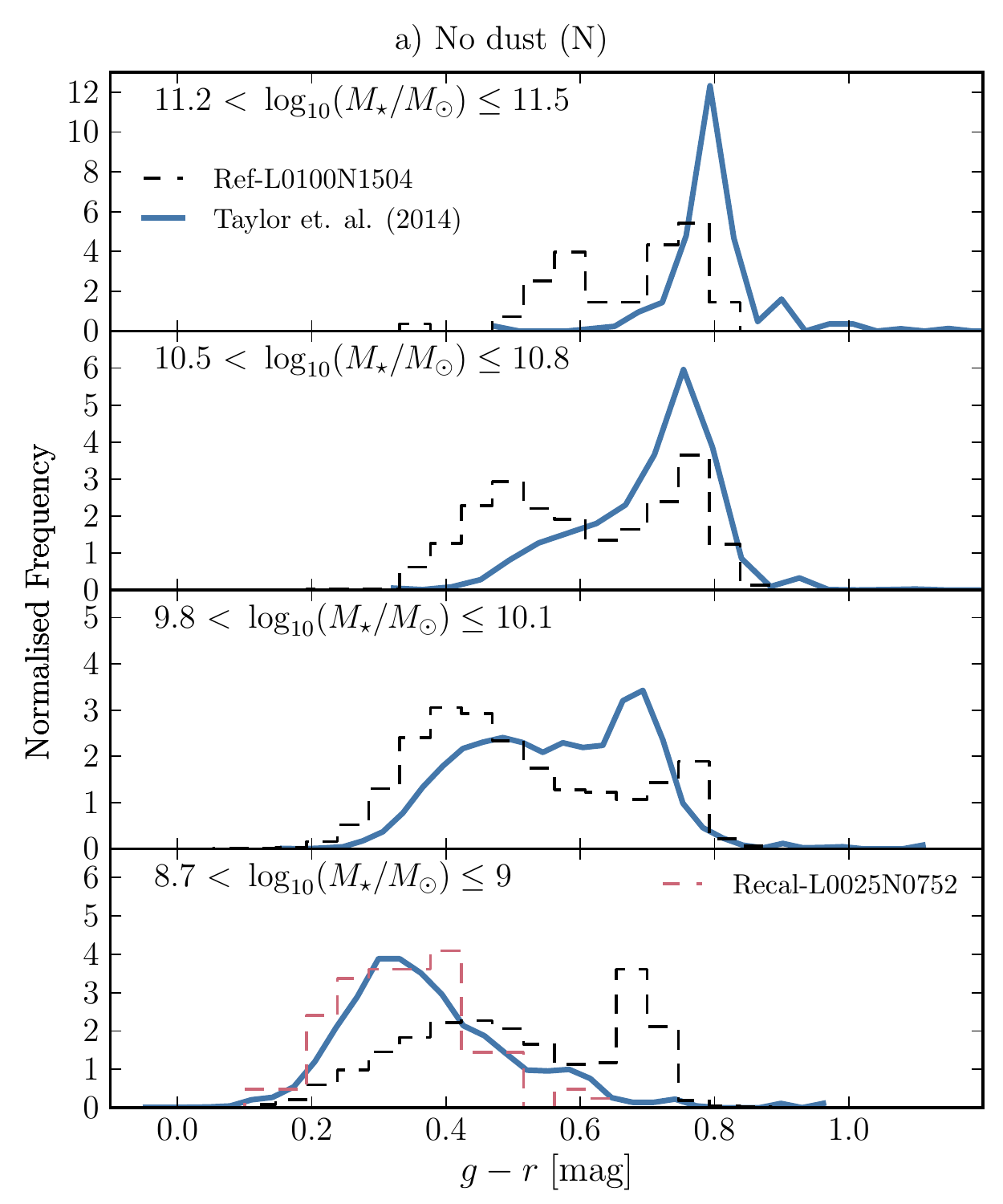}\\
  \includegraphics[width=\textwidth]{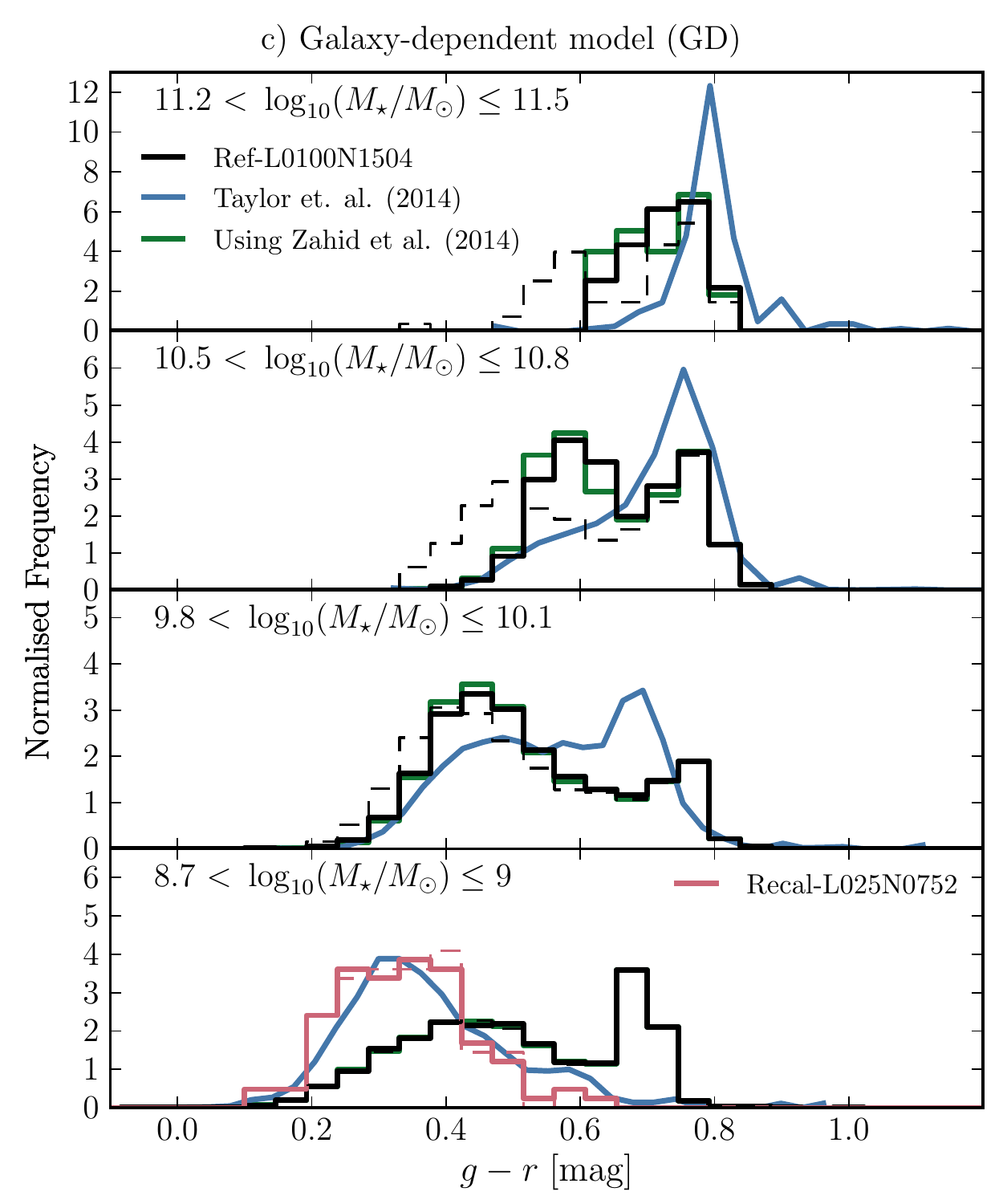}
    \end{minipage}
    \begin{minipage}{0.47\textwidth}
  \includegraphics[width=\textwidth]{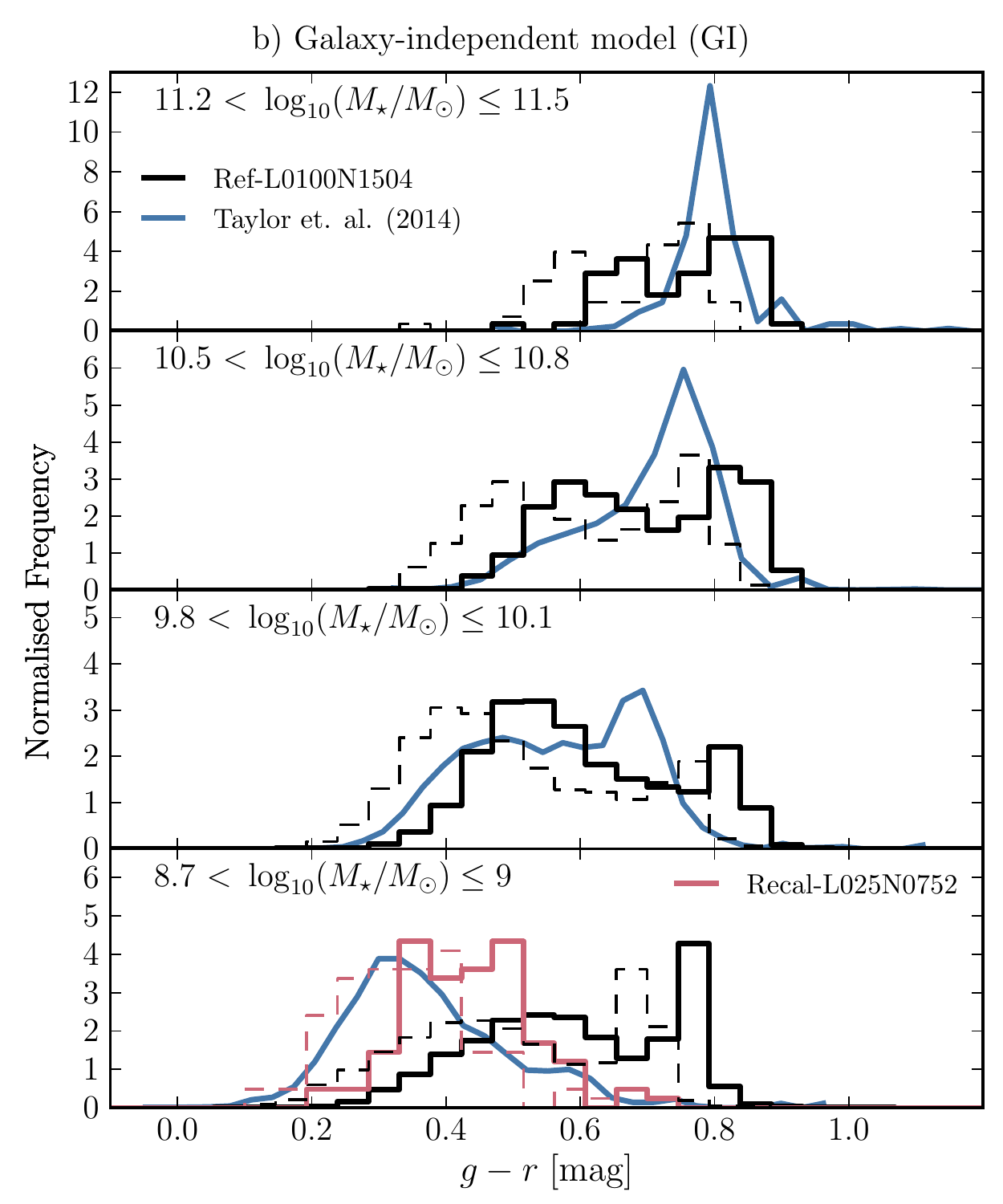}\\%figures/cmddevhist_b.pdf}\\
  \includegraphics[width=\textwidth]{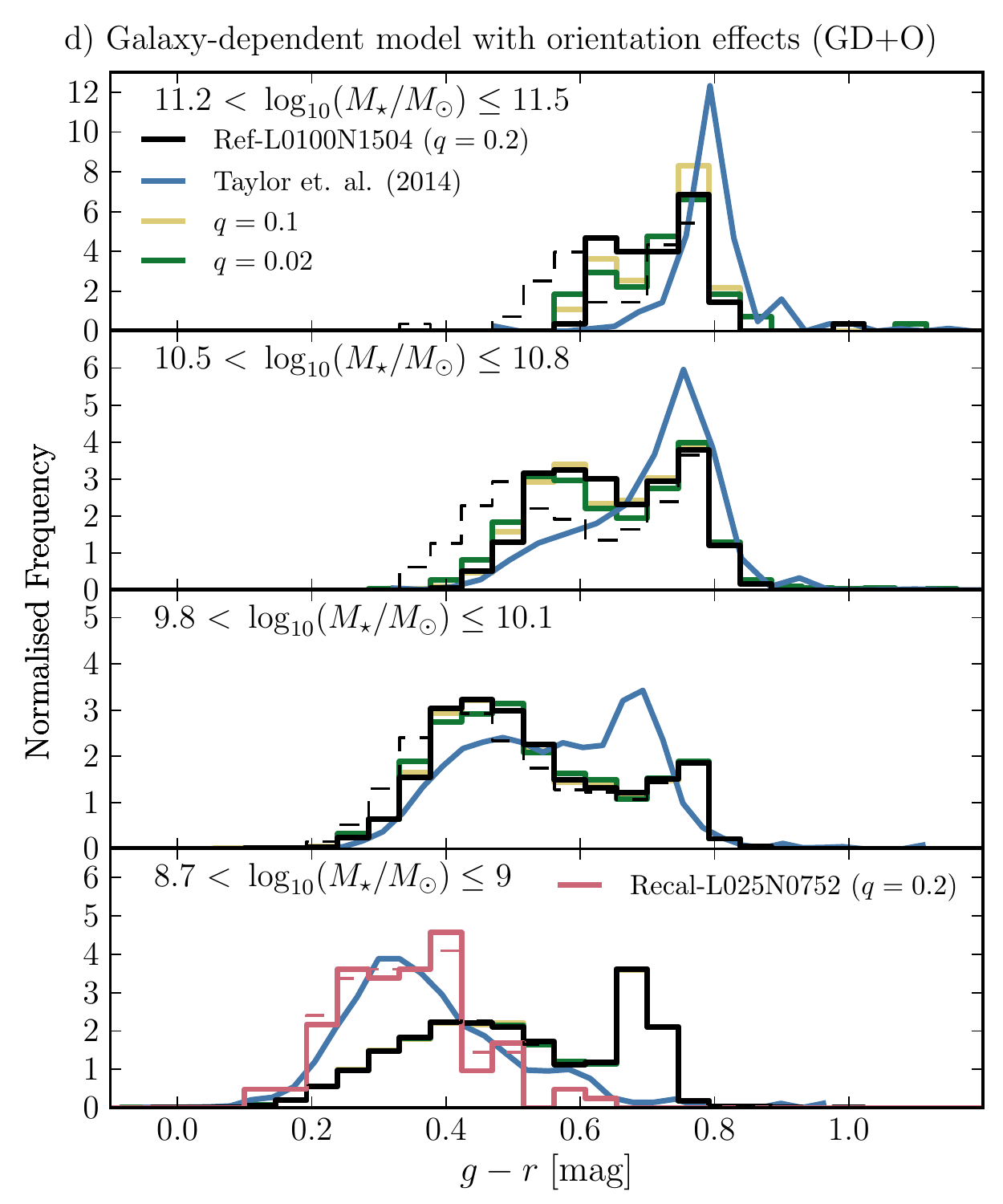}
    \end{minipage}
    \caption{Rest-frame $g-r$ colour distributions for \eagle{} galaxies at redshift $z = 0.1$, using 4 different
      models ({\em panels a-d}) for four non-contiguous ranges in stellar mass as indicated in the legend ({\em top to bottom}). 
      {\em Black lines} indicate
      the fiducial \ac{Ref-100} galaxy population while {\em red lines} indicate the higher-resolution \ac{Recal-25} simulation.
     {\em Dashed lines} denote the unobscured SED (model \modela{}); these are repeated in each panel for comparison.
    Models shown are: model \modela{} without dust (panel a, see \S\ref{modelE}), model \modelb{} with galaxy independent dust (panel b, see 
    \S~\ref{modelE-CF}), model \modelc{} where the dust obscuration
    depends on gas fraction and metallicity (panel c, \S~\ref{model:E-MCF}), and model \modeld{} that in addition takes into account orientation effects (panel d, \S~\ref{modelE-MCFO}). {\em Green} and {\em yellow lines} show model variations in panels c and d (see sections~\ref{model:E-MCF}--\ref{modelE-MCFO} for details). {\em Blue lines}
    represent observed galaxy colours for the volume-limited sample of {\sc gama} galaxies from \citet{Taylor14}. The figure shows the subtle quantitative effects that our different dust models have on the colour distributions of \eagle{}
      galaxies in various $M_\star$ regimes, as is discussed further in the text.
}
    \label{fig:cmddev}
\end{figure*}

\begin{figure*}
    \includegraphics[width=0.49\textwidth]{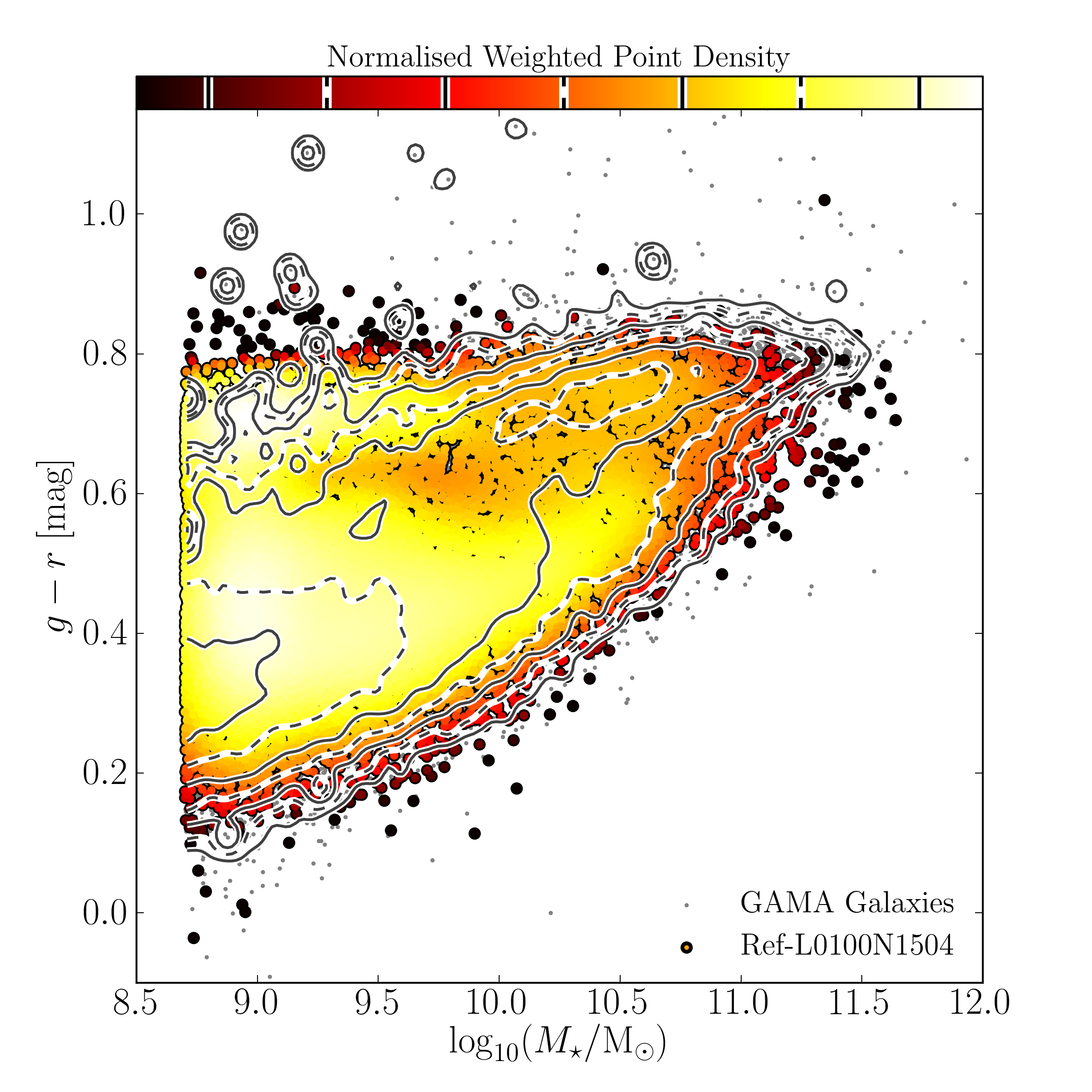}
    \includegraphics[width=0.49\textwidth]{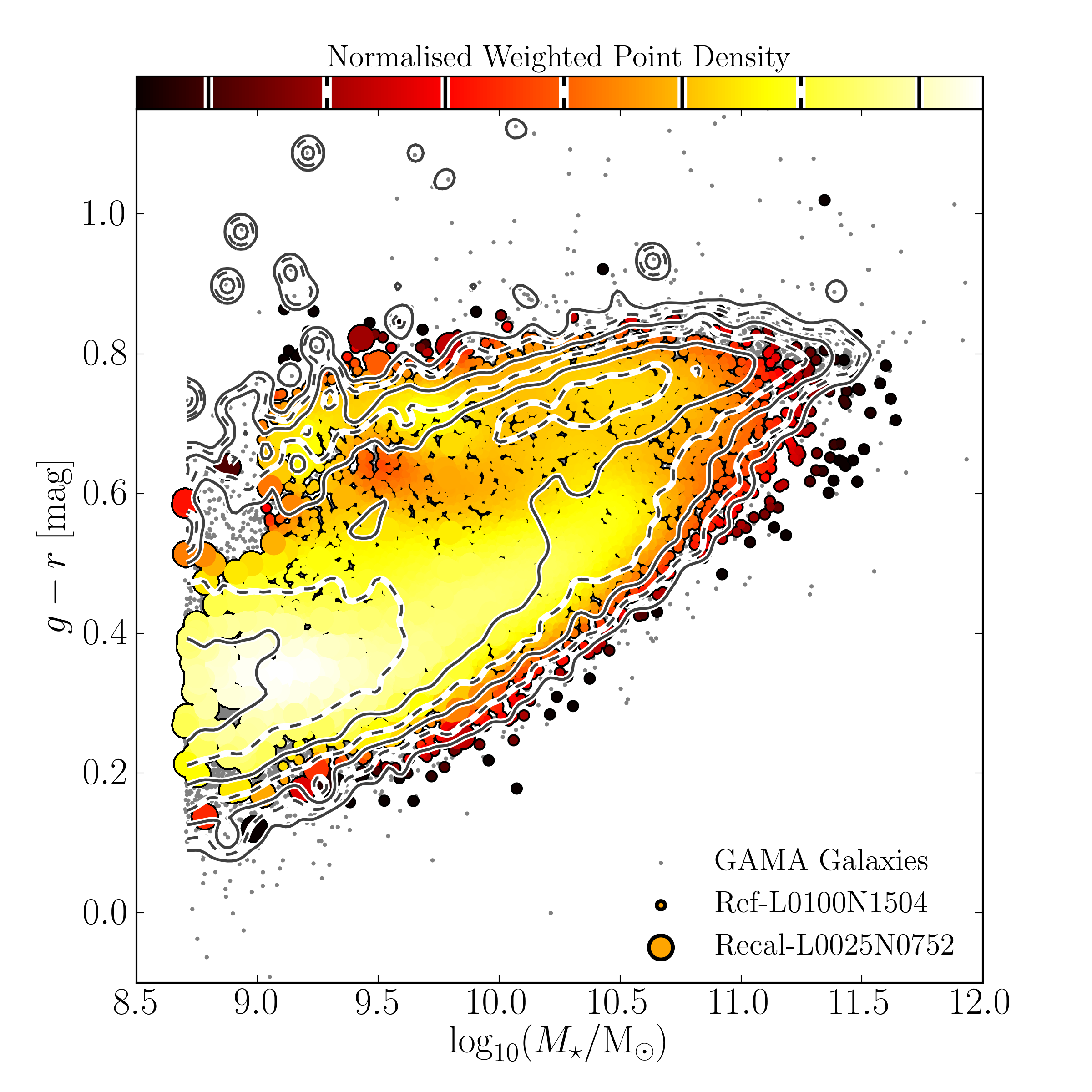}
         \caption{Rest-frame $g-r$ colour-stellar mass diagrams for {\sc
             eagle} galaxies using photometry and dust reddening from model \modeld{} at $z=0.1$ for simulation \ac{Ref-100} ({\em left panel}), and a composite set of \eagle{} galaxies obtained by combining the higher-resolution model \ac{Recal-25} for galaxies with $M_\star \lesssim 10^{9}$ M$_\odot$ and \ac{Ref-100} for $M_\star > 10^{9}$ M$_\odot$ ({\em right panel}), cross-fading the sampling probability of the two galaxy populations linearly in $\log_{10}(M_\star/{\rm M}_\odot)$ (see text for details). {\em Colours} represent the point density of \eagle{} galaxies of given $M_\star$ and $g-r$ colour (see text for details), with {\em black points} representing individual outlying \eagle{} galaxies. {\em Contours} represent the colour-$M_\star$ distribution for a volume-limited set of {\sc gama} galaxies from \citet{Taylor14}, with {\em grey points} representing individual outlying galaxies. The masses of observed galaxies are obtained through SED fitting, see \citet{Taylor14}. The colours are mapped to their equivalent contours in the colour bar. The colour bar covers 2 dex in point density with a contour spacing of 0.28 dex. A lack of numerical resolution makes lower-mass \eagle{} galaxies too red in the left panel. The transition of colours of observed galaxies, from the red-sequence at stellar masses above $M_\star\sim 10^{10.5}{\rm M}_\odot$ to the blue cloud at lower stellar masses, is reproduced in the simulation, although the blue cloud extends to slightly higher $M_\star$.
}
    \label{fig:cmd}
\end{figure*}

\label{sec:cmddev}
We contrast $g-r$ colours for \eagle{} galaxies in narrow (0.3 dex) bins of $M_\star$ for different models of dust absorption in Fig.~\ref{fig:cmddev} (panels a-d). In all panels, the blue line  corresponds to the observed distribution from the {\sc GAMA} survey \citep{Taylor14}. Different panels show models with no dust (model \modela{}, panel a), a dust model that is independent of galaxy type (model \modelb{}, panel b), a model in which dust opacity depends on metallicity and gas fraction (model \modelc{}, panel c), and finally a model that in addition accounts for orientation effects (model \modeld{}, panel d). All models are shown as histograms, normalised to have unit integral. Models including dust are plotted as solid histograms, while the dashed histograms represent model \modela{} in all panels.
\ac{Ref-100} distributions for the fiducial dust models are plotted in black, whereas \ac{Recal-25} distributions are plotted (for the lowest mass bin) in red. Model variations are also plotted for \ac{Ref-100}, with a model using the observed mass-metallicity relation in panel c shown in green (see section~\ref{model:E-MCF}) and models with alternative $q$ values in panel d shown in green and yellow (see section~\ref{modelE-MCFO}).

The observed $g-r$ colours shift from very red massive galaxies (usually termed the \lq red sequence\rq), to a broader distribution in colours for $10^{9.8}\lesssim M_\star/{\rm M}_\odot\lesssim 10^{10.8}$, and finally a blue population (usually termed \lq blue cloud\rq) at lower stellar masses. There is little evidence for a strong \lq bimodality\rq\ in observed colours
even though the data is often interpreted that way. Such an interpretation is perhaps due to the bimodality seen at a fixed optical magnitude, where blue galaxies are pushed into higher luminosity bins.

Before comparing the models to the data, we investigate the effects of dust modelling, going from high- to low-mass galaxies (top to bottom rows in panels a-d). The most massive galaxies in the \eagle{} \ac{Ref-100} simulation ($M_\star/{\rm M}_\odot \gtrsim 10^{11.2}$, top rows) have a relatively extended intrinsic colour distribution (model \modela{}). Including a model with dust reddening independent of galaxy properties incorrectly reddens the reddest galaxies even more (model \modelb{}) but taking into account the relatively low cold-gas masses of these galaxies returns the colours to close to their intrinsic values (model \modelc{}). The tail of bluer massive galaxies is significantly affected by dust, yielding a mono-modal distribution in the highest-mass bin. Including orientation effects (model \modeld{}) gives a slightly broader colour distribution for the fiducial axial ratio value of $q=0.2$. 

A similar trend is noticeable for galaxies in the second most massive
bin ($10^{10.5}\lesssim M_\star/{\rm M}_\odot\lesssim 10^{11.2}$,
second row from the top). Though the fiducial value of $q=0.2$ for
\modeld{} produces a similar distribution to \modelc{} in this bin,
varying $q$ values appears to have the strongest effect here. Smaller
$q$ values show a more pronounced bimodality, as the majority of
galaxies are subject to less reddening, while a minority of \lq{}edge-on\rq{} galaxies are heavily reddened to $g-r$ colours $\gtrsim 0.8$.

The third bin ($10^{9.8}\lesssim M_\star/{\rm M}_\odot\lesssim 10^{10.1}$, third row from the top) also behaves similarly: the intrinsically bluest galaxies get reddened slightly more than the intrinsically red galaxies with scatter due to orientation having a negligible effect. 

Finally, the colour distribution for galaxies in the least massive bin ($10^{8.7}\lesssim M_\star/{\rm M}_\odot\lesssim 10^{9}$, bottom row)
is also shown for the \ac{Recal-25} simulation. There is a large difference in colours between \ac{Ref-100} and \ac{Recal-25} for the least massive galaxies, which is predominately a resolution effect: at the resolution of \ac{Ref-100} the star formation rates in these low-mass galaxies is underestimated (S15) which makes the simulated galaxies too red. This striking resolution dependence is not surprising. In \ac{Ref-100}, galaxies of mass $M_\star\sim 10^{9}{\rm M}_\odot$ are represented by only $\sim 10^3$ star particles, and for a typical cold gas fraction of 10 per cent, by only 100 star-forming gas particles.  We demonstrate in Appendix \ref{sec:conv} that red and blue sequence colours for galaxies across the $10^{9.8}\lesssim M_\star/{\rm M}_\odot\lesssim 10^{10.1}$ range\footnote{A larger mass range is used in Appendix~\ref{sec:conv} than in the third row of Fig.~\ref{fig:cmddev} so that \ac{Ref-25} and \ac{Recal-25} are sufficiently sampled.} are quite similar in \ac{Ref-100} and \ac{Recal-25} - which gives us confidence that \ac{Recal-25} gives numerically converged answers for the bottom row of Fig.~\ref{fig:cmddev}. However, the different environments probed by the \ac{Ref-100} and \ac{Recal-25} models also contribute to the difference in colours, in particular the strength of the red sequence, because the larger volume contains a population of satellite galaxies in massive halos. This is also shown in Appendix \ref{sec:conv} and discussed further in section \ref{sec:disc}.
Taking into account dust obscuration and orientation effects has little effect on the colours in \ac{Recal-25} for these low-mass galaxies, with model \modeld{} and \modela{} yielding nearly identical colour distributions.

We now turn to comparing the colours of \eagle{} galaxies to the data, going from top (most massive) to bottom (least massive) bins in stellar mass and focusing on model \modeld{}, Fig.~\ref{fig:cmddev}d). At the massive end, the observed red sequence galaxies are about 0.05 mag redder in the data than in \eagle{}. As the optical colours of old ($\gtrsim 10$ Gyr) stellar populations are dominated by metallicity effects \citep{Bell69}, this small colour difference is attributable to SSP metallicities. The $M_\star-Z_\star$ relation for \eagle{} galaxies (S15) is seen to lie slightly below (by less than 0.1 dex) observational data in galaxies with $M_\star/{\rm M}_\odot > 10^{11}$, resulting in a slightly bluer red sequence colour. The data also has a tail to even redder colours not present in \eagle{}. In contrast, the most massive \eagle{} galaxies have a tail to {\em bluer} colours resulting from recent star formation. It could be that such star formation is shielded more effectively in the data ({\em i.e.} the value of $\hat\tau_{\rm bc}$ used is too low), or alternatively that our AGN feedback scheme does not quite suppress star formation sufficiently. The higher than observed gas fractions for galaxy clusters in \eagle{} (S15) could also contribute to the enhanced SFR of some simulated BCGs.

The red sequence of galaxies with $10^{10.5}\lesssim M_\star/{\rm M}_\odot\lesssim 10^{10.8}$ is very similar in the data and the simulation, but in \eagle{} there are significantly more blue galaxies. The blue cloud starts to appear in the data for galaxies with $10^{9.8}\lesssim M_\star/{\rm M}_\odot\lesssim 10^{10.1}$, and its colour is very similar in \eagle{}. However, in \eagle{} the blue peak is stronger and the red peak occurs at a slightly redder colours ($g-r = 0.75$ compared to the observed value of 0.7). Using smaller values of $q$ in \modeld{} does not improve agreement with observation here.
Dust reddening and orientation effects already play little role in setting the colours of \eagle{} galaxies in this mass bin. Finally, in the lowest mass bin, $10^{8.7}\lesssim M_\star/{\rm M}_\odot\lesssim 10^{9}$, there is excellent agreement in the colour distributions of simulation and data and once more our dust reddening models are unimportant in setting \eagle{} colours.

This level of agreement between galaxy colours in the simulation and the data is 
encouraging. By including metallicity and orientation effects in our dust treatment, we prevent the significant colour shift seen in the simple \modelb{} model. The validity of our dust model is discussed further in section \ref{sec:disc}. Despite the good agreement, there are some clear discrepancies between \eagle{} and the {\sc gama} colour distributions. These can be seen in the widths and relative strengths of red and blue populations. The latter discrepancy reflects the finding of S15 that the transition from actively star-forming to passive galaxies occurs at slightly (by a factor of $\sim 2$) too high mass in \eagle{}.

The dependence of galaxy colours on stellar mass is further illustrated in Fig.~\ref{fig:cmd}, where the number density of galaxies in \eagle{} with given rest-frame $g-r$ colour (computed using model \modeld{}) and stellar mass is compared to a volume-limited sample of {\sc gama} galaxies taken from \citet{Taylor14}. Results are plotted down to stellar masses of $10^{8.7}$M$_\odot$, below which volume corrections due to the influence of line-of-sight structure become increasingly uncertain in the data \citep{Taylor14}. The colour bar shows the point density of \eagle{} galaxies and how these map to the \citet{Taylor14} contours.

The left panel shows the galaxies taken from simulation \ac{Ref-100}. The simulation reproduces the trend seen in the data from galaxies being red above a stellar mass of $M_\star \sim 10^{10.5}{\rm M}_\odot$ to being predominantly blue below that. However, as also seen in the previous figure, there is a population of red ($g-r\sim 0.7$) low-mass ($M_\star\sim 10^9{\rm M}_\odot$) galaxies in \eagle{} that is not seen in the data. These galaxies are modelled using only $\sim 1000$ star particles; the right panel of Fig.~\ref{fig:cmd} therefore uses the higher resolution simulation \ac{Recal-25} for galaxies below $10^9{\rm M}_\odot$ and \ac{Ref-100} above $10^9{\rm M}_\odot$, cross-fading one simulation into the other linearly in $\log(M_\star)$\footnote{The sampling frequency of \ac{Recal-25}
galaxies in this plot are weighted a factor of 64 higher than
\ac{Ref-100}
galaxies to account for the smaller volume and a further factor of 2 lower to account for the boosted number counts in \ac{Recal-25} caused by poor sampling of large scale power in the smaller volume. This means that for masses $< 10^9{\rm M}_\odot$ each \ac{Recal-25} galaxy contributes a factor of 32 more to the point densities in Fig.~\ref{fig:cmd} than a 
\ac{Ref-100} galaxy at mass $>10^9{\rm M}_\odot$.}. With this we aim to show the colour-mass distribution for a larger range of well-resolved galaxies ($\gtrsim 1000$ star particles), while avoiding a discontinuity that renders the overall distribution less clear. 

Combining these two resolutions, the colours of \eagle{} galaxies at given $M_\star$ track the data from the {\sc gama} galaxies \citep{Taylor14} well. Both display a red sequence of massive galaxies which becomes redder with increasing stellar mass, and $g-r\sim 0.7$ at $M_\star\sim 10^{10.5}{\rm M}_\odot$. The simulation also reproduces the width of that sequence, albeit with a shallower slope.  A blue cloud of galaxies appears both in \eagle{} and {\sc gama} below $M_\star\sim 10^{10.5}{\rm M}_\odot$, with $g-r\sim 0.45$ at $M_\star=10^{10.5}{\rm M}_\odot$. At decreasing stellar mass, the location of the blue cloud becomes bluer, reaching $g-r\sim 0.35$ at $M_\star=10^9\,{\rm M}_\odot$. Overall we find that \eagle{} reproduces the mean trends in galaxy colours well. Though the eradication of the faint red sequence in this sample is at least partly due to improved sampling, it also comes about because these galaxies are much less abundant in the higher resolution \ac{Recal-25} simulation (further discussion of the origin of this faint red population can be found in Section \ref{sec:disc} and Appendix \ref{sec:conv}).

In addition to the mean location of galaxy colours, there are outliers in both data and simulation. The {\sc gama} data display a scatter to extremely red colours ($g-r>1$) at all stellar masses only seen for one high-mass outlier in \eagle{}. There is also a scattering of galaxies $\sim 0.1$~mag bluer than the main locus in {\sc gama} that appear in \eagle{} as well. Finally, \eagle{} has some very massive, relatively blue galaxies ($M_\star\sim 10^{11.5}{\rm M}_\odot$, $g-r\sim 0.6$); although there are such galaxies in {\sc gama} as well, they are more numerous in \eagle{}, as is more easily seen in Fig.~\ref{fig:cmddev}. We suggested before that these either imply too little dust reddening in star forming regions in \eagle{}, or simply that some of these massive \eagle{} galaxies are undergoing too much star formation despite the inclusion of AGN feedback.

\subsection{Luminosity functions and colour-magnitude diagram}
\begin{figure*}
  \centering
  \includegraphics[width=1.05\textwidth]{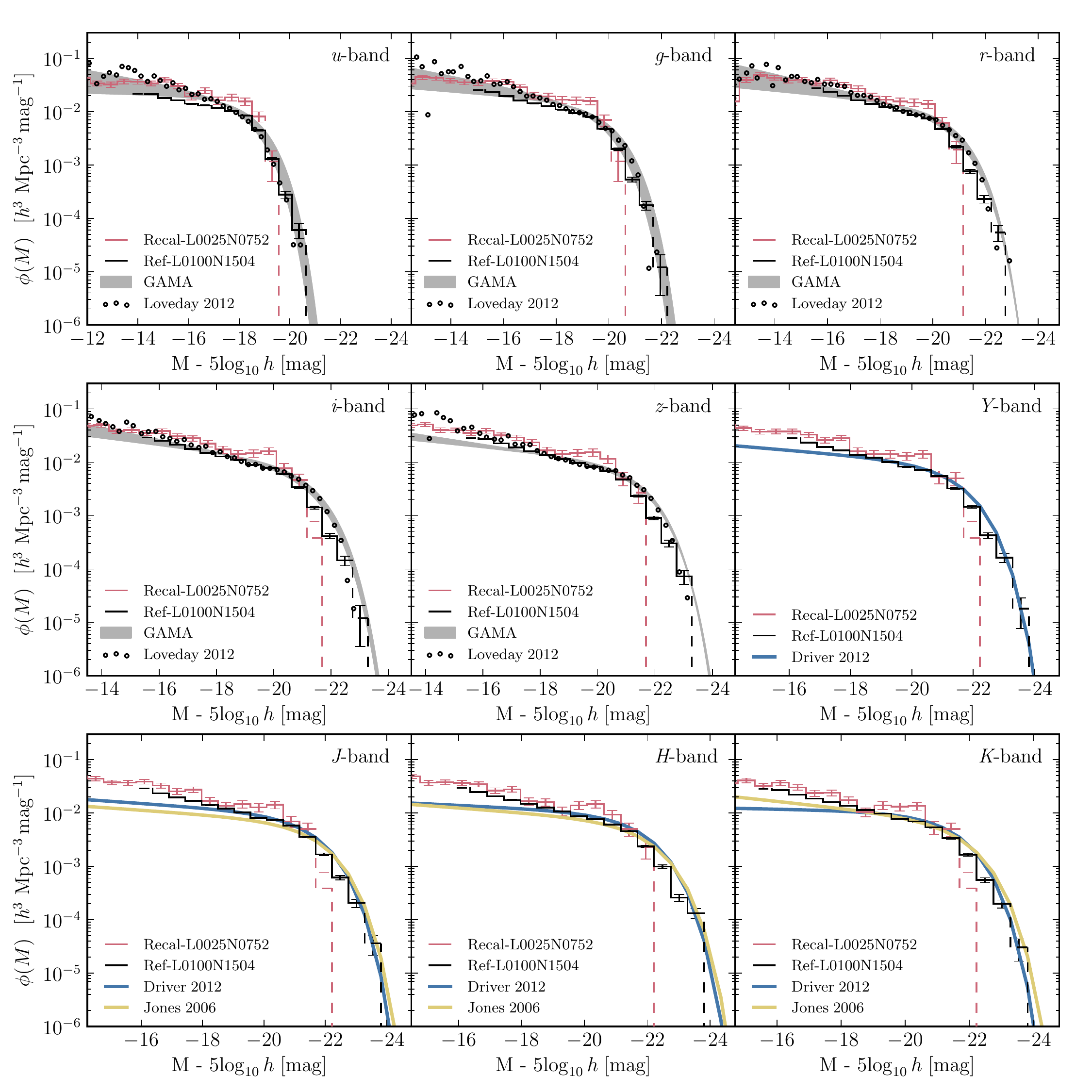}
  \caption{Luminosity functions in each band of the $ugrizYHJK$
    photometric system plotted for the fiducial 100 Mpc simulation (\ac{Ref-100}) in black
    and the re-calibrated high resolution 25 Mpc simulation (\ac{Recal-25}) in red, both  at
    redshift $z = 0.1$, using the \modeld{} dust model. The \ac{Ref-100} function is plotted down to
    the faintest magnitude bin at which most galaxies are represented by \textgreater
    100 star particles. The simulation results from $ugriz$ SDSS
      bands are compared to the Petrosian GAMA survey luminosity functions
      taken from \citet{Loveday12}, $k$-corrected to rest-frame magnitudes
      and plotted as empty circles. The region bound by the Schechter fits
      to the \citet{Driver12} and \citet{Loveday12} luminosity functions is
      shaded in grey. As these two luminosity functions are measured
      using Kron and Petrosian magnitudes respectively, the grey area indicates the
      difference due to aperture definition. Schechter fits for UKIDSS band luminosity functions
      are taken from \gama{} (Driver et. al. 2012) and from 6DF+2MASS \citep{HeathJones06} where
      available. All magnitudes are in the AB system. Error bars reflect Poisson errors
      with dashed lines indicating bins
      containing $<10$ simulation galaxies. The \eagle{} luminosity functions
      are similar to the observed fits across the spectral range,
      with some discrepancies discussed in Section \ref{sec:results}.}
    \label{fig:lfs}
\end{figure*}

\begin{table*}
   \footnotesize
 \centering
 \begin{minipage}{175mm}
   \centering
   \begin{tabular}{@{}llccc@{}}
     \hline
     Model & Description &  \multicolumn{3}{c|}{$r$-band}\\
      & & $\phi_\ast $ [$h^3$ cMpc$^{-3}$]& $\alpha$ 
      & $M_\star - 5\log_{10}(h)$ [mag] \\
     \hline
     \modela{} & No dust &$7.0^{+0.7}_{-0.5} \times 10^{-3}$ &
     $-1.26^{+0.02}_{-0.01}$ & $-21.2^{+0.2}_{-0.2}$\\
     \modelb{} & Galaxy-independent dust model & $7.6^{+1.0}_{-0.6} \times 10^{-3}$&
     $-1.23^{+0.04}_{-0.02}$& $-20.7^{+0.2}_{-0.1}$ \\
     \modelc{} & $Z$ and $M_{\rm ISM}$ dependent & $8.2^{+1.0}_{-1.0} \times
     10^{-3}$& $-1.23^{+0.05}_{-0.02}$ & $-21.0^{+0.3}_{-0.3}$\\
     \modeld{} & \modelc{} with orientation dependence &$8.3^{+1.3}_{-0.9} \times 10^{-3}$& $-1.23^{+0.04}_{-0.02}$&
     $-20.9^{+0.2}_{-0.2}$\\
     \hline
     Data & \citet{Loveday12}  & $9{\pm 0.7}\times
     10^{-3}$& $-1.26{\pm 0.01}$
     & $-20.7{\pm 0.03}$\\
     \hline

\end{tabular}
   \caption{Best-fitting Schechter function (Eq.~\ref{eq:Schechter}) parameters for \eagle{} AB-magnitude luminosity functions in the $r$-band for different dust models in simulation \ac{Ref-100} at redshift $z = 0.1$, and the observed luminosity function from \citet{Loveday12}. The \eagle{} and observed luminosity functions are all fit over the magnitude range $-23.2 < r < -14.2$, $1\sigma$ errors on the best-fit parameters were computed using jackknife sampling.}
\label{tab:schec}
\end{minipage}
\end{table*}

\label{sec:cmd}
Luminosity functions for model \modeld{} in rest-frame $ugrizYJHK$ broad-band filters are plotted using absolute AB magnitudes in Fig.~\ref{fig:lfs}. The simulations \ac{Ref-100} and \ac{Recal-25} at redshift $z=0.1$ are shown with Poisson error bars as solid black and red histograms respectively, becoming dashed when there are fewer than 10 galaxies per $\sim$0.6 mag bin. For \ac{Recal-25} the bins are correlated, as can be seen for example in the $u$-band for bins $M-5\log(h)=-17.5$ to $-18.5$, due to poor sampling of large-scale modes in the small volume.  There is generally good agreement between the two runs, with \ac{Recal-25} typically less than a factor of two ($0.3$~dex) higher at the faint end, and by much less for the redder bands. Note that this higher-resolution simulation does not sample the exponential cut-off at high luminosities because of its small volume. Differences in resolution are most noticeable in bluer colours, particularly in $u$. As discussed before, in small galaxies the stellar feedback events driving outflows are poorly sampled and the star forming components are poorly resolved. As a result star formation rates and thus intrinsic colours are subject to considerable resolution effects. We see that the higher-resolution simulation yields higher star formation rates and hence bluer colours. We again note that this is not just a resolution issue: the \ac{Ref-100} volume contains a population of faint red quenched satellites of massive galaxies, which are simply not present in the much smaller \ac{Recal-25} volume, as discussed further in Appendix~\ref{sec:conv}. LFs in longer wavelength bands are consistent in shape between the two simulations (despite the small volume simulation being noisier).

Observed luminosity functions from \citet{Loveday12} are plotted in each of the $ugriz$ bands, which we fit with a single \citet{Schechter76} function,
\begin{equation}
{1\over L_\star}\,{dn\over dL} = \phi_\star\,\left({L\over L_\star}\right)^\alpha\,\exp(-L/L_\star)\,{dL\over L_\star}\,.
\label{eq:Schechter}
\end{equation}
Single Schechter function fits are also taken from \citet{Driver12} and \citet{HeathJones06}. For the $ugriz$ bands these are shown as grey shaded regions which are bounded by the fits to the observed luminosity function of \citet{Loveday12} and \citet{Driver12}, both based on data from {\sc gama}. The differences between these observed luminosity functions result from the use of Kron and Petrosian magnitudes, respectively. The thickness of the grey band is thus a measure of how these different aperture choices affect the Schechter fit. For the $YJHK$ band, we plot published Schechter fits, which are based on {\sc ukidss} data.

There appears to be some discrepancy between Schechter fits from the observational papers and directly observed luminosity functions in the optical, as can be seen by comparing the \citet{Loveday12} data with the Schechter fits.
In particular, the data points appear systematically higher than the Schechter fit at the faint end and below the fit at the bright end. This is most visible in the $z$-band where the shaded region is narrowest. This could be a consequence of 
intermediate magnitude bins dominating the fit as this is where observational errors are minimal. It also shows that the single Schechter function is not a good fit to the observational data. In particular, it is unable to represent
the observations at the faint end accurately \citep[e.g.][]{Loveday98}. The \citet{Driver12} and \citet{HeathJones06} fits agree reasonably well, except for the faint-end slope in the $K$-band. The single Schechter fits are used here as simple indicators of the shape, position and normalisation of the observed luminosity functions, but clearly their exact location depends on details of how galaxies are identified in the data, and possibly on the range and assumed errors used in the fitting procedure. We compare the parameters of Schechter fits to \eagle{} luminosity functions to observational fits in Table~\ref{tab:schec}. 

From Table~\ref{tab:schec} we see that the dust treatment has little effect on the shape of the $r$-band luminosity function. The effect of including dust using the \modelb{} model makes the knee position, $M_\ast$, 0.5 mag fainter magnitudes and increases the normalisation, $\phi_\ast$. Scaling dust absorption by galaxy properties in \modelc{} serves to increase $\phi_\ast$ further, and makes $M_\ast$ brighter by 0.3 mag. The \modelc{} and \modeld{} model luminosity functions provide Schechter fits that agree with the observational values within the errors. The faint-end slope, $\alpha$, shows very little variation between all dust models, and all are consistent with the observational fit within the errors.

Comparing the \modeld{} \eagle{} luminosity functions to the data in Fig. \ref{fig:cmd} shows a striking overall consistency from the UV to the NIR bands. The deviations are mostly of the same order as differences in fits to the published luminosity functions of different authors. The agreement is particularly good in the optical bands $ugriz$, where \eagle{} tends to fall mostly inside the grey band that represents the dependence of the luminosity function on the choice of aperture. The excellent agreement over such a wide range of colours suggests that \eagle{} forms the correct number of galaxies of a given stellar mass and that those galaxies have realistic star formation histories and metallicities. 

Comparing blue bands ($u-g$) to redder bands ($J-K$) at the faint end, we notice that the \eagle{} \ac{Ref-100} luminosity functions tend to be slightly low in blue bands relative to the data, but high in the red bands. This is a consequence of \ac{Ref-100} producing slightly too many low-mass galaxies (S15) which have too low star formation rates \citep{Furlong14}. Resolution also plays a role: we plot galaxies with more than 100 star particles, where we know that the stellar feedback events generating outflows and star formation rates at the faint end are poorly resolved. In addition, even fainter galaxies with high star formation rates cannot scatter into the faint-end bins since we impose a cut in mass and not in magnitude. 

The \eagle{} luminosity function tends to drop below the observations at the \lq knee\rq\  ($L_\star$) in the Schechter function, particularly in the bands red-ward of $r$. This is consistent with a slight underestimate in the masses of more massive \eagle{} galaxies, as seen in the mass function plotted in S15. 

The $JHK$ bands also appear to have generally somewhat steeper faint-end slopes (parameter $\alpha$ in Eq.\ref{eq:Schechter}) in \eagle{} than the Schechter fits to the data. However, the data itself also shows a upturn at the faint end relative to the observed Schechter fits \citep[circles in Fig. \ref{fig:lfs}, see also][]{Driver12}. Generally, the single Schechter function fit tends to underestimate the luminosity function at the faint end \citep{Loveday98}. This is more pronounced in the $JHK$ bands where the NIR sky is relatively bright \citep[e.g.][]{Sivanandam12}, leading to large uncertainties in the faint end data.

\subsection{The $g-r$ colour-magnitude diagram}
\begin{figure}
  \centering
  \includegraphics[width=0.98\columnwidth]{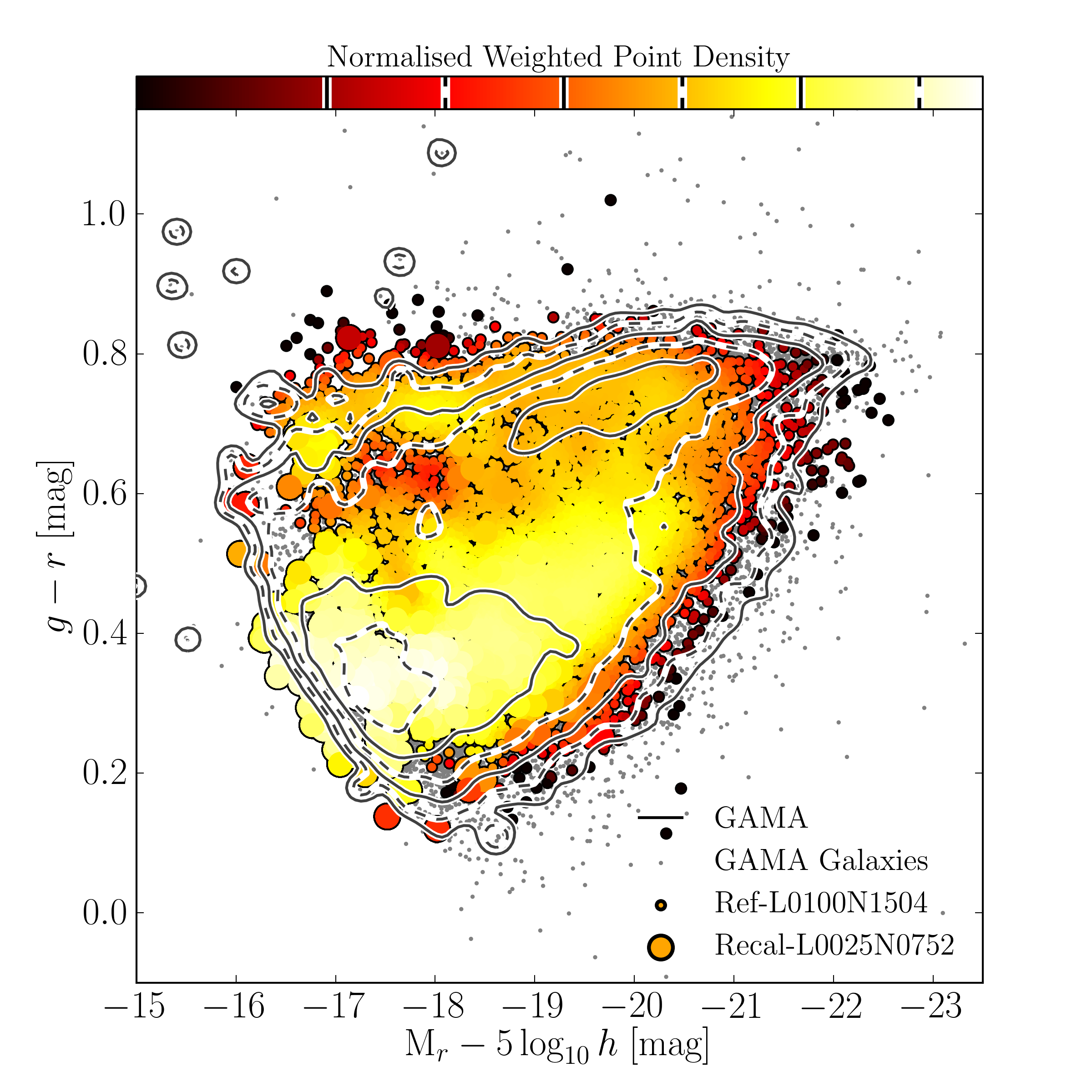}
    \caption{Rest-frame $g-r$ colour as a function of $r$-band absolute magnitude for \eagle{} galaxies (colours) compared to a volume-limited sample of {\sc gama} galaxies \citep[contour lines]{Taylor14}. contours and point shading is the same as in Fig \ref{fig:cmd}. \eagle{} photometry is obtained using the \modeld{} model. A composite \eagle{} galaxy population is used, consisting of galaxies from \ac{Ref-100} at $M_\star>10^9{\rm M}_\odot$ and \ac{Recal-25}
      at $M_\star<10^9{\rm M}_\odot$ as in the right panel of Fig \ref{fig:cmd}. There is excellent agreement between \eagle{} and {\sc gama} in the location and slope of the red-sequence ($g-r\approx 0.7$ at $M_r-5\log_{10} h=-20$), the appearance of a blue cloud of galaxies with $g-r\sim 0.45$ at that magnitude, which becomes increasingly blue ($g-r\sim 0.3$) for the fainter galaxies with $M_r-5\log h\sim -17.5$.      
      } 
    \label{fig:colmag}
\end{figure}

The colour-magnitude diagram of model \modeld{} for \eagle{} is plotted in Fig.~\ref{fig:colmag}. As in the right panel of Fig.~\ref{fig:cmd}, we combined faint galaxies from the higher-resolution simulation \ac{Recal-25} for galaxies with $M_\star<10^9{\rm M}_\odot$ with galaxies from simulation \ac{Ref-100} at higher masses.
As before, colours represent the number density of \eagle{} galaxies in this plane, whereas contours show the corresponding data from the volume-limited catalogue of {\sc gama} taken from 
\cite{Taylor14}. For both simulation and data, we only show galaxies with stellar mass $M_\star > 10^{8.7} {\rm M}_\odot$.

Fig.~\ref{fig:colmag} now contains only \lq observable\rq\ quantities for {\sc gama} galaxies\footnote{In practice the diagram still depends to a small extent on the applied cut in stellar mass at the faint end and on the choice of aperture to measure magnitudes.}, and in particular does not require any SED fitting. The overall agreement between \eagle{} and the data is very good: the location and slope of the red-sequence is in good agreement, the transition to the blue cloud at $M_r-5\log(h)\sim 20$ is reproduced, and the blue $g-r\sim 0.3$ colours of the fainter blue galaxies are also reproduced. At the imposed mass cut of $M_\star > 10^{8.7} {\rm M}_\odot$, the colours and magnitudes of \eagle{} galaxies agree well with the data. 
The bright blue galaxies that appear in \eagle{} become slightly more discrepant with the data in this plot than in Fig.~\ref{fig:cmd}, due to the fact that bluer galaxies generally possess brighter $r$-band magnitudes for the same
$M_\star$.

The level of agreement between simulation and data in the colour-magnitude diagram shown in Fig.~\ref{fig:colmag} is remarkable and suggests that these \eagle{} simulations provide a relatively realistic population of galaxies at low redshift, and that the modelling of emission and dust obscuration in model \modeld{} works well.

\section{Discussion}
\label{sec:disc}

The \eagle{} simulations were calibrated to reproduce the local GSMF and galaxy sizes by appropriate choice of the parameters in the subgrid model for feedback.
As stellar mass is closely linked to NIR luminosity, the consistency of GSMF proxies such as the
$K$-band luminosity function shown in Fig. \ref{fig:lfs} is not surprising, at least at $z\sim 0$. 
However, consistency with the luminosity functions in other broad band filters is not automatic because stellar mass, star formation history, metallicity and dust obscuration all play a role.

In Section \ref{sec:results}, we focused on our \modelc{} and \modeld{} photometric models where dust absorption is approximated by a simple two component screen, with optical depths that vary with galactic gas content, metallicity and orientation (\modeld{} only). The colour distributions as a function of stellar mass and $r$-band magnitude for model \modeld{} in Figs.~\ref{fig:cmddev}, \ref{fig:cmd} and \ref{fig:colmag} show a level of agreement between simulated galaxies and observations that appears unprecedented for hydrodynamical simulations, and comparable to that achieved for semi-analytical models \citep[e.g.][]{Gonzalez09, Henriques14}. The \eagle{} luminosity functions also agree well with observations over a range in wavelengths from optical to NIR (Fig.~\ref{fig:lfs}). The relatively good agreement for number density, luminosity and colour, suggests that in \eagle{} each dark matter halo forms a galaxy with stellar mass, age and metallicity close to those inferred from observation. The similar level of agreement in observed colour-magnitude space also rules out a potential circularity resulting from using the same photometric model to infer stellar mass from observational data as is used in the simulations. Even though the overall level of agreement is good between \eagle{} and the data, there are discrepancies.

There is an excess of bright ($M_r-5\log_{10} h\lesssim -20$) blue ($g-r\lesssim 0.6$) \eagle{} galaxies relative to data, apparent in e.g. Fig.~\ref{fig:colmag}.
Such an excess is seen in all our photometric models (see Fig.~\ref{fig:cmddev}), but is least apparent for model \modeld{} where the recent star formation, that is the root cause of the blue colours, is most strongly obscured by dust. This parallels the findings of S15, that the fraction of passive \eagle{} galaxies is too low at the high-mass end, relative to observations. It may be that massive \eagle{} galaxies are too highly star forming, perhaps as a consequence of insufficient suppression of star formation by AGN. Whether the good agreement in the colour of the blue cloud, despite the underestimate of the median star formation rates at these masses (S15), suggests underestimated reddening is discussed further below.

The \eagle{} red sequence, at $g-r\sim 0.75$, is flatter than observed, both when plotted as a function of stellar mass (Fig.~\ref{fig:cmd}) and
when plotted as a function of absolute magnitude (Fig.~\ref{fig:colmag}). The flatter slope may be attributable to the dependence of stellar metallicity on galaxy mass, $Z_\star(M_\star)$; although stellar metallicities of \eagle{} galaxies agree well with the data at the massive end, they fall less rapidly with decreasing stellar mass compared to the observational data of \citet{Gallazzi05}, as shown in S15. Numerical resolution may play a role here, because the $Z_\star(M_\star)$ of the higher-resolution simulation \ac{Recal-25} does agree with the data; see S15.

There is an abundant population of red ($g-r\sim 0.7$), low-mass ($M_\star\sim 10^9{\rm M}_\odot$) galaxies in simulation \ac{Ref-100} that is not observed (see Fig.~\ref{fig:cmd}a). The comparison of simulations \ac{Ref-25} and \ac{Recal-25} in Appendix~B shows that this is at least partially due to a lack of numerical resolution. Indeed, star formation and outflows driven by feedback in these galaxies are poorly resolved and poorly sampled, leading to too low values of the specific star formation rate (and correspondingly too high passive fractions) and too high metallicities (see S15).  Appendix~\ref{sec:resmp} discusses the re-sampling technique we use in all models to try to mitigate poor sampling. Although this goes some way towards improving the modelling, it does not eliminate the discrepancy. Because the re-sampling is a post-processing step, it cannot help with the poor sampling of stellar feedback in these low-mass systems within the simulation. Related resolution problems are more intractable, and higher-resolution simulations are required to alleviate them.

Comparing simulations \ac{Ref-100} and \ac{Ref-25} that have identical numerical resolution (and the associated poor sampling of star formation in $M_\star\sim 10^9{\rm M}_\odot$ galaxies), yet differ in simulated volume size, allows us to isolate the effects of environment (see Appendix~\ref{sec:conv}). Although on average the colours of galaxies agree well between these simulations, the presence of faint red galaxies is much more pronounced in the larger volume. This is because many of these galaxies are satellites of more massive systems that are absent in the smaller volume. The fraction of satellites increases at lower stellar masses, and in the range $ 10^{8.7}{\rm M}_\odot < M_\star < 10^9{\rm M}_\odot$ comprises $\sim 46\%$ of the galaxy population in the \ac{Ref-100} simulation and $\sim 33\%$ in \ac{Ref-25}. Evidently, satellites contribute significantly to the colour distribution at low masses.  At present we cannot verify whether improved resolution will also reduce the suppression of star formation in, or decrease the metallicities of, small satellite galaxies, which would improve the colours of $M_\star\sim 10^9{\rm M}_\odot$ galaxies compared to data. We conclude  that the redder colours of low-mass galaxies in \eagle{} relative to data is at least partially a result of resolution, stemming from poor sampling of star formation and feedback. The improved agreement with the data that comes about
from using a composite sample of \ac{Recal-25} and \ac{Ref-100} galaxies relative to using the \ac{Ref-100} sample alone is
thus mainly due to improved numerical resolution, but also to the exclusion of red satellite galaxies that are not present in the smaller volume.

The level of agreement between \eagle{} colours and the data also depends on the realism of our dust reddening model. Fig. \ref{fig:cmddev} illustrates how dust reddening depends on the assumptions made in models \modela{} (no reddening) to model \modeld{} (gas metallicity, gas mass, and orientation-dependent reddening). Differences between these models are typically of order $\Delta (g-r)\sim 0.1$. A dust model that is independent of galaxy properties 
(such as \modelb{}) incorrectly reddens red galaxies. A reddening model that takes into account the gas mass (\modelc{}) resolves this inconsistency, with most of the remaining effects of reddening affecting blue bright galaxies. Overall, we find that the details of the dust treatments make relatively little difference to galaxy colours so that differences with observations are more likely due to the ages and metallicities of the stars rather than dust obscuration. This may be due to the relatively small effect of dust at redshifts $z \sim 0.1$. We leave the investigation of evolution of colours and luminosities in \eagle{} to a future work.

Taken at face value, the specific star formation rates of star-forming galaxies in \eagle{} are lower than inferred from observations by $\sim 0.2$~dex (S15). The fact that the colours of those same galaxies nevertheless agree with the data might imply that we underestimate dust reddening. Indeed, an underestimate of the gas fraction would lead to both an underestimate of the specific star formation rate and, at fixed metallicity, the attenuation. \citet{Lagos15} have shown that at $M_\ast \sim 10^{10} {\rm M}_\odot$ the median H$_2$ fraction in \eagle{} is about 0.2 dex lower than observed, and that this discrepancy goes away at higher masses and for the higher-resolution \ac{Recal-25} model.

Systematically lower attenuation for faint galaxies could thus be attributable to their low gas masses in the simulations. However, more complex models yield non-zero levels of attenuation even for very low gas surface densities \citep[e.g.][]{Boquien13}. The realism of mixed screen models as used here has been shown to break down when screens are optically thick 
\citep[e.g.][]{Disney89}. As dust optical depths are expected to be higher in blue bands, our dust prescription may be too crude to reproduce the data at higher levels of obscuration.  This could contribute to the bluer colours of massive galaxies in \eagle{}.

However, it is also possible that the levels of obscuration {\em are}
realistic, but that star formation rates are overestimated in the
data due to the absolute calibration of observed tracers of star
formation. The calibration of star formation rates from tracers rely on
assumptions about the intrinsic UV continuum (from population synthesis
modelling) and absorption at short wavelengths, as well as an assumed
form for the IMF \citep[e.g.][]{Kennicutt98}.  The cumulative build-up of stellar mass in \eagle{} is lower than observed by about 0.1~dex, whereas the star formation rate is lower than that observed by 0.2-0.4~dex, depending on redshift \citep{Furlong14}. This slight tension may suggest a small overestimate of the observationally inferred star formation rates. Estimating intrinsic properties from observables
of simulated galaxies (such as star formation rates) may help to clarify
these issues, see e.g. the recent study by \citet{Torrey14}. 

The dust model we developed here was designed to be as simple as possible, yet to avoid unrealistic levels of reddening. The model assigns 
a single value of reddening per galaxy without taking into account the non-uniform distributions of dust apart from that assigned to birth clouds. It is possible to make much more detailed estimates of reddening using 3D radiative transfer (RT) calculations \citep[e.g.][]{Baes05, Jonsson09}. We postpone comparisons of the current simple model to those obtained with the radiative transfer code {\sc skirt} \citep{Baes05} to future work.

\section{Summary \& Conclusions}
\label{sec:conclusion}

We have calculated broad band luminosities of simulated galaxies taken from the \eagle{} (Evolution and Assembly of GaLaxies and their Environments) suite of hydrodynamic simulations \citep[S15,][]{Crain15}, and compared them to observations of the redshift $z\sim 0.1$ galaxy population. The model uses simple stellar population modelling based on the {\sc galaxev} population synthesis models of \citet{bc03}. To reduce sampling noise arising from single young star particle in poorly resolved galaxies, we use the re-sampling procedure described in Appendix~\ref{sec:resmp}. In all models, galaxy luminosities are found by summing the particle luminosities within a 30~pkpc radius spherical aperture for consistency with previous analysis \citep[S15,][]{Furlong14}, which has been shown to mimic Petrosian apertures. Absolute magnitudes are presented in the AB system.

We compare and contrast three models of dust obscuration and to model \modela{} which neglects dust. Model \modelb{}, inspired by \cite{CF00}, includes contributions to the dust optical depth from the birth clouds of young stars and from a constant dust screen, with parameters that are independent of the galaxy properties. Applying a single diffuse dust correction to all galaxies incorrectly reddens ellipticals and we avoid this with model \modelc{} in which dust reddening depends on gas phase metallicity as well as gas mass. Finally, model \modeld{} uses a simple geometric model to account for orientation effects, which are however small.  

These simple models allow us to investigate the dependence of galaxy colours on stellar metallicities and ages, gas metallicities, and dust obscuration. Our main conclusions are as follows:

\begin{itemize}

\item The \modelb{} dust prescription which applies a reddening that is independent of galaxy properties, and was used by e.g. \citet{Torrey14}, excessively reddens the red-sequence population of galaxies. As a consequence, $g-r$ colours of massive ($M_\star\gtrsim 10^{10.5}{\rm M}_\odot$) \eagle{} galaxies are $\sim 0.1$~mag
  redder than observed, in spite of having ages and stellar metallicities that are similar ro those inferred(Fig~\ref{fig:cmddev}b). Scaling dust optical
  depths with cold gas mass and gas metallicity, as in model \modelc{}, is more realistic and improves agreement with observation (Fig~\ref{fig:cmddev}c and \ref{fig:cmddev}d).
  
\item The red sequence in \eagle{} is $\sim 0.1$~mag bluer in $g-r$ than observed for $M_\star\gtrsim 10^{11.2}{\rm M}_\odot$, and has a shallower dependence on stellar mass than observed (Fig.~\ref{fig:cmddev}). This is most likely a consequence of the dependence of colour on stellar metallicities.
  
\item The appearance of a faint red sequence in the
  \ac{Ref-100} simulation run ($M_\star < 10^{9.75}{\rm M}_\odot$, $0.6 < g-r < 0.8$, see left panel of Fig~\ref{fig:cmd}) that is not observed, is largely an
  effect of numerical resolution. Star formation and outflows are not well resolved in galaxies of such low mass.
  
\item A \lq blue cloud\rq\ of star forming galaxies appears in \eagle{} below $M_\star\ltsima 10^{10.5}{\rm M}_\odot$, with $g-r$ colour in agreement with the {\sc gama} data from \cite{Taylor14} (left panel of Fig~\ref{fig:cmd}). 

\item There is an excess of  bright ($M_r-5\log_{10} h\lesssim -20$) blue ($g-r\lesssim 0.6$) galaxies in \eagle{} relative to the data. This may be caused by an underestimate of the reddening in star forming regions, or an overestimate of the star formation rates in these massive galaxies due to insufficient suppression of star formation by AGN.
  
\item The $z=0.1$ galaxies taken from \eagle{} transition from mostly red ($g-r\sim 0.7$) above $M_\star\sim 10^{10.5}{\rm M}_\odot$ to mostly blue (ranging from $g-r\sim 0.5$ at $M_\star\sim 10^{10.5}{\rm M}_\odot$ becoming bluer with decreasing mass to $g-r\sim 0.35$ at $M_\star\sim 10^9{\rm M}_\odot$) at lower masses, follows the colours of {\sc gama} galaxies from \citet{Taylor14} (see Fig.\ref{fig:cmd}). However the blue cloud persists to higher than observed stellar masses, consistent with a similar trend in passive fractions shown in S15.

\item The $z=0.1$ galaxy luminosity functions constructed from the \eagle{}
  population agree well with data from UV to NIR bands, with differences of the order of the difference between using Kron and Petrosian magnitudes in the data (Fig.~\ref{fig:lfs}). This level of agreement is similar to the 
agreement between the \eagle{} and observed stellar mass functions. In particular, there is a slight underestimate in the number density of galaxies close to the knee of the Schechter fit, and the faint-end tends to be slightly steeper than observed in most bands. We note, however, that the faint-end of the luminosity function is uncertain, especially in NIR bands, and single Schechter fits tend to underestimate the faint-end slope \citep{Loveday98}. Good agreement was not surprising in the NIR where luminosities are dominated by stellar mass, whereas the good agreement in other bands suggests that the star formation histories and metal enrichment in \eagle{} galaxies are relatively realistic.
  
\item The $z=0.1$ $g-r$ colour versus $M_r$ magnitude diagram for galaxies with $M_\star\gtrsim 10^9{\rm M}_\odot$ yields a level of agreement with data that is comparable to that of current semi-analytic models \citep[Fig.~\ref{fig:colmag};][]{Gonzalez09, Henriques14}. The similar
colour distributions of \modela{} and \modeld{} photometries (Fig.~\ref{fig:cmddev}d) suggests that the dust model plays only a minor role in this agreement. This further attests to the relatively realistic evolution of the \eagle{} galaxy population. 

\end{itemize}

The general agreement in the colour and luminosity of \eagle{} galaxies and observed galaxies suggests that the simulated galaxies have similar star formation histories, metal enrichment processes, and current star formation rates as observed galaxies. This makes the \eagle{} suite well-suited to investigate the physical processes that shape galaxies through cosmic time.

\section*{Acknowledgements}
The authors would like to extend gratitude to Ned Taylor and John Loveday for the kind provision of observational datasets for comparison in this work.
The authors would also like to thank Alessandro Bressan, St\`{e}phane Charlot and Claudia Maraston for their invaluable insight into the calibration and application of stellar population models.
The authors also thank the GAMA Team for the use of GAMA survey data
products in advance of its public release from the GAMA website http://www.gama-survey.org/.
This work was supported by the Science and Technology Facilities Council [grant number ST/F001166/1], 
by the Interuniversity Attraction Poles Programme initiated by the
Belgian Science Policy Office ([AP P7/08 CHARM], by ERC grant agreement 278594 - GasAroundGalaxies,
and used the DiRAC Data Centric system at Durham University, operated by the Institute
for Computational Cosmology on behalf of the STFC DiRAC HPC Facility
(www.dirac.ac.uk). This equipment was funded by BIS National
E-infrastructure capital grant ST/K00042X/1, STFC capital grant
ST/H008519/1, and STFC DiRAC Operations grant ST/K003267/1 and Durham
University. DiRAC is part of the National E-Infrastructure. RAC is a
Royal Society University Research Fellow.
The data used in the work is available through collaboration with the authors.

\bibliographystyle{mn2e}

\bibliography{references}

\begin{thebibliography}{110}
\expandafter\ifx\csname natexlab\endcsname\relax\def\natexlab#1{#1}\fi

\bibitem[{{Allende Prieto} {et~al.}(2001){Allende Prieto}, {Lambert}, \&
  {Asplund}}]{Allende01}
{Allende Prieto} C., {Lambert} D.~L., {Asplund} M., 2001, \apjl, 556, L63

\bibitem[{{Asplund} {et~al.}(2004){Asplund}, {Grevesse}, {Sauval}, {Allende
  Prieto}, \& {Kiselman}}]{Asplund04}
{Asplund} M., {Grevesse} N., {Sauval} A.~J., {Allende Prieto} C., {Kiselman}
  D., 2004, \aap, 417, 751

\bibitem[{{Baes} {et~al.}(2005){Baes}, {Dejonghe}, \& {Davies}}]{Baes05}
{Baes} M., {Dejonghe} H., {Davies} J.~I., 2005, in American Institute of
  Physics Conference Series, Vol. 761, The Spectral Energy Distributions of
  Gas-Rich Galaxies: Confronting Models with Data, {Popescu} C.~C., {Tuffs}
  R.~J., eds., pp. 27--38

\bibitem[{{Baldry} {et~al.}(2014){Baldry}, {Alpaslan}, {Bauer},
  {Bland-Hawthorn}, {Brough}, {Cluver}, {Croom}, {Davies}, {Driver},
  {Gunawardhana}, {Holwerda}, {Hopkins}, {Kelvin}, {Liske},
  {L{\'o}pez-S{\'a}nchez}, {Loveday}, {Norberg}, {Peacock}, {Robotham}, \&
  {Taylor}}]{Baldry14}
{Baldry} I.~K., {Alpaslan} M., {Bauer} A.~E., {et~al.}, 2014, \mnras, 441, 2440

\bibitem[{{Baldry} {et~al.}(2004){Baldry}, {Glazebrook}, {Brinkmann},
  {Ivezi{\'c}}, {Lupton}, {Nichol}, \& {Szalay}}]{baldry04}
{Baldry} I.~K., {Glazebrook} K., {Brinkmann} J., {et~al.}, 2004, \apj, 600, 681

\bibitem[{{Baldry} {et~al.}(2010){Baldry}, {Robotham}, {Hill}, {Driver},
  {Liske}, {Norberg}, {Bamford}, {Hopkins}, {Loveday}, {Peacock}, {Cameron},
  {Croom}, {Cross}, {Doyle}, {Dye}, {Frenk}, {Jones}, {van Kampen}, {Kelvin},
  {Nichol}, {Parkinson}, {Popescu}, {Prescott}, {Sharp}, {Sutherland},
  {Thomas}, \& {Tuffs}}]{Baldry10}
{Baldry} I.~K., {Robotham} A.~S.~G., {Hill} D.~T., {et~al.}, 2010, \mnras, 404,
  86

\bibitem[{Bell \& Rodgers(1969)}]{Bell69}
Bell R., Rodgers A., 1969, Monthly Notices of the Royal Astronomical Society,
  142, 161

\bibitem[{{Benson} {et~al.}(2003){Benson}, {Bower}, {Frenk}, {Lacey}, {Baugh},
  \& {Cole}}]{Benson03}
{Benson} A.~J., {Bower} R.~G., {Frenk} C.~S., {et~al.}, 2003, \apj, 599, 38

\bibitem[{{Bernardi} {et~al.}(2013){Bernardi}, {Meert}, {Sheth}, {Vikram},
  {Huertas-Company}, {Mei}, \& {Shankar}}]{Bernardi13}
{Bernardi} M., {Meert} A., {Sheth} R.~K., {et~al.}, 2013, \mnras, 436, 697

\bibitem[{{Booth} \& {Schaye}(2009)}]{Booth09}
{Booth} C.~M., {Schaye} J., 2009, \mnras, 398, 53

\bibitem[{{Boquien} {et~al.}(2013){Boquien}, {Boselli}, {Buat}, {Baes},
  {Bendo}, {Boissier}, {Ciesla}, {Cooray}, {Cortese}, {Eales}, {Koda},
  {Lebouteiller}, {de Looze}, {Smith}, {Spinoglio}, \& {Wilson}}]{Boquien13}
{Boquien} M., {Boselli} A., {Buat} V., {et~al.}, 2013, \aap, 554, A14

\bibitem[{{Bower} {et~al.}(2006){Bower}, {Benson}, {Malbon}, {Helly}, {Frenk},
  {Baugh}, {Cole}, \& {Lacey}}]{Bower06}
{Bower} R.~G., {Benson} A.~J., {Malbon} R., {et~al.}, 2006, \mnras, 370, 645

\bibitem[{{Bressan} {et~al.}(1993){Bressan}, {Fagotto}, {Bertelli}, \&
  {Chiosi}}]{Bressan93}
{Bressan} A., {Fagotto} F., {Bertelli} G., {Chiosi} C., 1993, \aaps, 100, 647

\bibitem[{Bruzual \& Charlot(2003)}]{bc03}
Bruzual G., Charlot S., 2003, Monthly Notices of the Royal Astronomical
  Society, 344, 1000

\bibitem[{{Chabrier}(2003)}]{Chabrier03}
{Chabrier} G., 2003, \pasp, 115, 763

\bibitem[{{Charlot} \& {Fall}(2000)}]{CF00}
{Charlot} S., {Fall} S.~M., 2000, \apj, 539, 718

\bibitem[{{Cole} {et~al.}(2000){Cole}, {Lacey}, {Baugh}, \& {Frenk}}]{Cole00}
{Cole} S., {Lacey} C.~G., {Baugh} C.~M., {Frenk} C.~S., 2000, \mnras, 319, 168

\bibitem[{{Conroy} {et~al.}(2009){Conroy}, {Gunn}, \& {White}}]{Conroy09}
{Conroy} C., {Gunn} J.~E., {White} M., 2009, \apj, 699, 486

\bibitem[{{Crain} {et~al.}(2015){Crain}, {Schaye}, {Bower}, {Furlong},
  {Schaller}, {Theuns}, {Dalla Vecchia}, {Frenk}, {McCarthy}, {Helly},
  {Jenkins}, {Rosas-Guevara}, {White}, \& {Trayford}}]{Crain15}
{Crain} R.~A., {Schaye} J., {Bower} R.~G., {et~al.}, 2015, ArXiv e-prints,
  astro-ph/1501.01311

\bibitem[{{Crain} {et~al.}(2009){Crain}, {Theuns}, {Dalla Vecchia}, {Eke},
  {Frenk}, {Jenkins}, {Kay}, {Peacock}, {Pearce}, {Schaye}, {Springel},
  {Thomas}, {White}, \& {Wiersma}}]{Crain09}
{Crain} R.~A., {Theuns} T., {Dalla Vecchia} C., {et~al.}, 2009, \mnras, 399,
  1773

\bibitem[{{Creasey} {et~al.}(2013){Creasey}, {Theuns}, \& {Bower}}]{Creasey13}
{Creasey} P., {Theuns} T., {Bower} R.~G., 2013, \mnras, 429, 1922

\bibitem[{{Creasey} {et~al.}(2015){Creasey}, {Theuns}, \& {Bower}}]{Creasey15}
---, 2015, \mnras, 446, 2125

\bibitem[{{Croton} {et~al.}(2006){Croton}, {Springel}, {White}, {De Lucia},
  {Frenk}, {Gao}, {Jenkins}, {Kauffmann}, {Navarro}, \& {Yoshida}}]{Croton06}
{Croton} D.~J., {Springel} V., {White} S.~D.~M., {et~al.}, 2006, \mnras, 365,
  11

\bibitem[{{Cullen} \& {Dehnen}(2010)}]{Cullen10}
{Cullen} L., {Dehnen} W., 2010, \mnras, 408, 669

\bibitem[{{da Cunha} {et~al.}(2008){da Cunha}, {Charlot}, \&
  {Elbaz}}]{daCunha08}
{da Cunha} E., {Charlot} S., {Elbaz} D., 2008, \mnras, 388, 1595

\bibitem[{{Dalla Vecchia} \& {Schaye}(2012)}]{DallaVecchia12}
{Dalla Vecchia} C., {Schaye} J., 2012, \mnras, 426, 140

\bibitem[{{Davis} {et~al.}(1985){Davis}, {Efstathiou}, {Frenk}, \&
  {White}}]{Davis85}
{Davis} M., {Efstathiou} G., {Frenk} C.~S., {White} S.~D.~M., 1985, \apj, 292,
  371

\bibitem[{{Dekel} \& {Silk}(1986)}]{Dekel86}
{Dekel} A., {Silk} J., 1986, \apj, 303, 39

\bibitem[{{Disney} {et~al.}(1989){Disney}, {Davies}, \& {Phillipps}}]{Disney89}
{Disney} M., {Davies} J., {Phillipps} S., 1989, \mnras, 239, 939

\bibitem[{{Doi} {et~al.}(2010){Doi}, {Tanaka}, {Fukugita}, {Gunn}, {Yasuda},
  {Ivezi{\'c}}, {Brinkmann}, {de Haars}, {Kleinman}, {Krzesinski}, \& {French
  Leger}}]{SDSSfilters}
{Doi} M., {Tanaka} M., {Fukugita} M., {et~al.}, 2010, \aj, 139, 1628

\bibitem[{{Dolag} {et~al.}(2009){Dolag}, {Borgani}, {Murante}, \&
  {Springel}}]{Dolag09}
{Dolag} K., {Borgani} S., {Murante} G., {Springel} V., 2009, \mnras, 399, 497

\bibitem[{{Driver}(2012)}]{Driver12}
{Driver} S.~P. {\em et al.}., 2012, \mnras, 427, 3244

\bibitem[{{Driver} {et~al.}(2011){Driver}, {Hill}, {Kelvin}, {Robotham},
  {Liske}, {Norberg}, {Baldry}, {Bamford}, {Hopkins}, {Loveday}, {Peacock},
  {Andrae}, {Bland-Hawthorn}, {Brough}, {Brown}, {Cameron}, {Ching}, {Colless},
  {Conselice}, {Croom}, {Cross}, {de Propris}, {Dye}, {Drinkwater}, {Ellis},
  {Graham}, {Grootes}, {Gunawardhana}, {Jones}, {van Kampen}, {Maraston},
  {Nichol}, {Parkinson}, {Phillipps}, {Pimbblet}, {Popescu}, {Prescott},
  {Roseboom}, {Sadler}, {Sansom}, {Sharp}, {Smith}, {Taylor}, {Thomas},
  {Tuffs}, {Wijesinghe}, {Dunne}, {Frenk}, {Jarvis}, {Madore}, {Meyer},
  {Seibert}, {Staveley-Smith}, {Sutherland}, \& {Warren}}]{Driver11}
{Driver} S.~P., {Hill} D.~T., {Kelvin} L.~S., {et~al.}, 2011, \mnras, 413, 971

\bibitem[{{Driver} {et~al.}(2009){Driver}, {Norberg}, {Baldry}, {Bamford},
  {Hopkins}, {Liske}, {Loveday}, {Peacock}, {Hill}, {Kelvin}, {Robotham},
  {Cross}, {Parkinson}, {Prescott}, {Conselice}, {Dunne}, {Brough}, {Jones},
  {Sharp}, {van Kampen}, {Oliver}, {Roseboom}, {Bland-Hawthorn}, {Croom},
  {Ellis}, {Cameron}, {Cole}, {Frenk}, {Couch}, {Graham}, {Proctor}, {De
  Propris}, {Doyle}, {Edmondson}, {Nichol}, {Thomas}, {Eales}, {Jarvis},
  {Kuijken}, {Lahav}, {Madore}, {Seibert}, {Meyer}, {Staveley-Smith},
  {Phillipps}, {Popescu}, {Sansom}, {Sutherland}, {Tuffs}, \&
  {Warren}}]{Driver09}
{Driver} S.~P., {Norberg} P., {Baldry} I.~K., {et~al.}, 2009, Astronomy and
  Geophysics, 50, 12

\bibitem[{{Durier} \& {Dalla Vecchia}(2012)}]{Durier12}
{Durier} F., {Dalla Vecchia} C., 2012, \mnras, 419, 465

\bibitem[{{Efstathiou}(1992)}]{Efstathiou92}
{Efstathiou} G., 1992, \mnras, 256, 43P

\bibitem[{{Fischera} {et~al.}(2003){Fischera}, {Dopita}, \&
  {Sutherland}}]{Fischera03}
{Fischera} J., {Dopita} M.~A., {Sutherland} R.~S., 2003, \apjl, 599, L21

\bibitem[{{Furlong} {et~al.}(2014){Furlong}, {Bower}, {Theuns}, {Schaye},
  {Crain}, {Schaller}, {Dalla Vecchia}, {Frenk}, {McCarthy}, {Helly},
  {Jenkins}, \& {Rosas-Guevara}}]{Furlong14}
{Furlong} M., {Bower} R.~G., {Theuns} T., {et~al.}, 2014, ArXiv e-prints,
  astro-ph/1410.3485

\bibitem[{{Gallazzi} {et~al.}(2005){Gallazzi}, {Charlot}, {Brinchmann},
  {White}, \& {Tremonti}}]{Gallazzi05}
{Gallazzi} A., {Charlot} S., {Brinchmann} J., {White} S.~D.~M., {Tremonti}
  C.~A., 2005, \mnras, 362, 41

\bibitem[{{Genel} {et~al.}(2014){Genel}, {Vogelsberger}, {Springel}, {Sijacki},
  {Nelson}, {Snyder}, {Rodriguez-Gomez}, {Torrey}, \& {Hernquist}}]{Genel14}
{Genel} S., {Vogelsberger} M., {Springel} V., {et~al.}, 2014, \mnras, 445, 175

\bibitem[{{Gonz{\'a}lez} {et~al.}(2009){Gonz{\'a}lez}, {Lacey}, {Baugh},
  {Frenk}, \& {Benson}}]{Gonzalez09}
{Gonz{\'a}lez} J.~E., {Lacey} C.~G., {Baugh} C.~M., {Frenk} C.~S., {Benson}
  A.~J., 2009, \mnras, 397, 1254

\bibitem[{Gonzalez-Perez {et~al.}(2014)Gonzalez-Perez, Lacey, Baugh, Lagos,
  Helly, Campbell, \& Mitchell}]{GonzalezPerez14}
Gonzalez-Perez V., Lacey C., Baugh C., {et~al.}, 2014, Monthly Notices of the
  Royal Astronomical Society, 439, 264

\bibitem[{{Gonzalez-Perez} {et~al.}(2014){Gonzalez-Perez}, {Lacey}, {Baugh},
  {Lagos}, {Helly}, {Campbell}, \& {Mitchell}}]{Gonzalez14}
{Gonzalez-Perez} V., {Lacey} C.~G., {Baugh} C.~M., {et~al.}, 2014, \mnras, 439,
  264

\bibitem[{{Grootes} {et~al.}(2013){Grootes}, {Tuffs}, {Popescu}, {Pastrav},
  {Andrae}, {Gunawardhana}, {Kelvin}, {Liske}, {Seibert}, {Taylor}, {Graham},
  {Cooray}, {Dariush}, {De Zotti}, {Driver}, {Dunne}, {Gomez}, {Hopkins},
  {Hopwood}, {Jarvis}, {Loveday}, {Maddox}, {Madore}, {Micha{\l}owski},
  {Norberg}, {Parkinson}, {Prescott}, {Robotham}, {Smith}, {Thomas}, \&
  {Valiante}}]{Grootes13}
{Grootes} M.~W., {Tuffs} R.~J., {Popescu} C.~C., {et~al.}, 2013, \apj, 766, 59

\bibitem[{{Haardt} \& {Madau}(2001)}]{Haardt01}
{Haardt} F., {Madau} P., 2001, in Clusters of Galaxies and the High Redshift
  Universe Observed in X-rays, {Neumann} D.~M., {Tran} J.~T.~V., eds.

\bibitem[{{Henriques} {et~al.}(2014){Henriques}, {White}, {Thomas}, {Angulo},
  {Guo}, {Lemson}, {Springel}, \& {Overzier}}]{Henriques14}
{Henriques} B., {White} S., {Thomas} P., {et~al.}, 2014, ArXiv e-prints,
  astro-ph/1410.0365

\bibitem[{{Henriques} {et~al.}(2013){Henriques}, {White}, {Thomas}, {Angulo},
  {Guo}, {Lemson}, \& {Springel}}]{Henriques13}
{Henriques} B.~M.~B., {White} S.~D.~M., {Thomas} P.~A., {et~al.}, 2013, \mnras,
  431, 3373

\bibitem[{{Hewett} {et~al.}(2006){Hewett}, {Warren}, {Leggett}, \&
  {Hodgkin}}]{UKIRTfilters}
{Hewett} P.~C., {Warren} S.~J., {Leggett} S.~K., {Hodgkin} S.~T., 2006, \mnras,
  367, 454

\bibitem[{{Hill} {et~al.}(2011){Hill}, {Kelvin}, {Driver}, {Robotham},
  {Cameron}, {Cross}, {Andrae}, {Baldry}, {Bamford}, {Bland-Hawthorn},
  {Brough}, {Conselice}, {Dye}, {Hopkins}, {Liske}, {Loveday}, {Norberg},
  {Peacock}, {Croom}, {Frenk}, {Graham}, {Jones}, {Kuijken}, {Madore},
  {Nichol}, {Parkinson}, {Phillipps}, {Pimbblet}, {Popescu}, {Prescott},
  {Seibert}, {Sharp}, {Sutherland}, {Thomas}, {Tuffs}, \& {van
  Kampen}}]{Hill11}
{Hill} D.~T., {Kelvin} L.~S., {Driver} S.~P., {et~al.}, 2011, \mnras, 412, 765

\bibitem[{{Hopkins}(2013)}]{Hopkins13}
{Hopkins} P.~F., 2013, \mnras, 428, 2840

\bibitem[{{Hopkins} {et~al.}(2011){Hopkins}, {Quataert}, \&
  {Murray}}]{Hopkins11}
{Hopkins} P.~F., {Quataert} E., {Murray} N., 2011, \mnras, 417, 950

\bibitem[{{Inoue}(2012)}]{Inoue12}
{Inoue} A.~K., 2012, ArXiv e-prints, astro-ph/1202.2932

\bibitem[{{Jackson} {et~al.}(2010){Jackson}, {Bryan}, {Mao}, \&
  {Li}}]{Jackson10}
{Jackson} N., {Bryan} S.~E., {Mao} S., {Li} C., 2010, \mnras, 403, 826

\bibitem[{{Jones} {et~al.}(2006){Jones}, {Peterson}, {Colless}, \&
  {Saunders}}]{HeathJones06}
{Jones} D.~H., {Peterson} B.~A., {Colless} M., {Saunders} W., 2006, \mnras,
  369, 25

\bibitem[{{Jonsson} {et~al.}(2009){Jonsson}, {Groves}, \& {Cox}}]{Jonsson09}
{Jonsson} P., {Groves} B., {Cox} T.~J., 2009, ArXiv e-prints,
  astro-ph/0906.2156

\bibitem[{{Kennicutt}(1998)}]{Kennicutt98}
{Kennicutt} Jr. R.~C., 1998, \araa, 36, 189

\bibitem[{{Kewley} \& {Ellison}(2008)}]{Kewley08}
{Kewley} L.~J., {Ellison} S.~L., 2008, \apj, 681, 1183

\bibitem[{{Lacey} \& {Cole}(1994)}]{Lacey94}
{Lacey} C., {Cole} S., 1994, \mnras, 271, 676

\bibitem[{{Lagos} {et~al.}(2015){Lagos}, {Crain}, {Schaye}, {Furlong}, {Frenk},
  {Bower}, {Schaller}, {Theuns}, {Trayford}, {Bahe}, \& {Dalla
  Vecchia}}]{Lagos15}
{Lagos} C.~d.~P., {Crain} R.~A., {Schaye} J., {et~al.}, 2015, ArXiv e-prints,
  astro-ph/1503.04807

\bibitem[{{Larson}(1974)}]{Larson74}
{Larson} R.~B., 1974, \mnras, 169, 229

\bibitem[{{Lawrence}(2007)}]{Lawrence07}
{Lawrence} A.~{\em et al.}., 2007, \mnras, 379, 1599

\bibitem[{{Leitherer} {et~al.}(1999){Leitherer}, {Schaerer}, {Goldader},
  {Delgado}, {Robert}, {Kune}, {de Mello}, {Devost}, \& {Heckman}}]{sb99}
{Leitherer} C., {Schaerer} D., {Goldader} J.~D., {et~al.}, 1999, \apjs, 123, 3

\bibitem[{{Loveday}(1998)}]{Loveday98}
{Loveday} J., 1998, ArXiv Astrophysics e-prints, astro-ph/astro-ph/9805255

\bibitem[{{Loveday} {et~al.}(2012){Loveday}, {Norberg}, {Baldry}, {Driver},
  {Hopkins}, {Peacock}, {Bamford}, {Liske}, {Bland-Hawthorn}, {Brough},
  {Brown}, {Cameron}, {Conselice}, {Croom}, {Frenk}, {Gunawardhana}, {Hill},
  {Jones}, {Kelvin}, {Kuijken}, {Nichol}, {Parkinson}, {Phillipps}, {Pimbblet},
  {Popescu}, {Prescott}, {Robotham}, {Sharp}, {Sutherland}, {Taylor}, {Thomas},
  {Tuffs}, {van Kampen}, \& {Wijesinghe}}]{Loveday12}
{Loveday} J., {Norberg} P., {Baldry} I.~K., {et~al.}, 2012, \mnras, 420, 1239

\bibitem[{Maraston(2005)}]{m05}
Maraston C., 2005, Monthly Notices of the Royal Astronomical Society, 362, 799

\bibitem[{{Martizzi} {et~al.}(2014){Martizzi}, {Faucher-Gigu{\`e}re}, \&
  {Quataert}}]{Martizzi14}
{Martizzi} D., {Faucher-Gigu{\`e}re} C.-A., {Quataert} E., 2014, ArXiv
  e-prints, astro-ph/1409.4425

\bibitem[{{McMillan}(2011)}]{McMillan11}
{McMillan} P.~J., 2011, \mnras, 414, 2446

\bibitem[{Mo {et~al.}(2010)Mo, van~den Bosch, \& White}]{Mo10}
Mo H., van~den Bosch F., White S., 2010, Galaxy Formation and Evolution, Galaxy
  Formation and Evolution. Cambridge University Press

\bibitem[{{Okamoto} {et~al.}(2008){Okamoto}, {Gao}, \& {Theuns}}]{Okamoto08}
{Okamoto} T., {Gao} L., {Theuns} T., 2008, \mnras, 390, 920

\bibitem[{{Oke}(1974)}]{Oke74}
{Oke} J.~B., 1974, \apjs, 27, 21

\bibitem[{{Oppenheimer} {et~al.}(2010){Oppenheimer}, {Dav{\'e}}, {Kere{\v s}},
  {Fardal}, {Katz}, {Kollmeier}, \& {Weinberg}}]{Oppenheimer10}
{Oppenheimer} B.~D., {Dav{\'e}} R., {Kere{\v s}} D., {et~al.}, 2010, \mnras,
  406, 2325

\bibitem[{{Planck Collaboration} {et~al.}(2014){Planck Collaboration}, {Ade},
  {Aghanim}, {Armitage-Caplan}, {Arnaud}, {Ashdown}, {Atrio-Barandela},
  {Aumont}, {Baccigalupi}, {Banday}, \& et~al.}]{Planck}
{Planck Collaboration}, {Ade} P.~A.~R., {Aghanim} N., {et~al.}, 2014, \aap,
  571, A16

\bibitem[{{Porter} {et~al.}(2014){Porter}, {Somerville}, {Primack}, \&
  {Johansson}}]{Porter14}
{Porter} L.~A., {Somerville} R.~S., {Primack} J.~R., {Johansson} P.~H., 2014,
  \mnras, 444, 942

\bibitem[{{Price}(2008)}]{Price08}
{Price} D.~J., 2008, Journal of Computational Physics, 227, 10040

\bibitem[{{Puchwein} \& {Springel}(2013)}]{Puchwein13}
{Puchwein} E., {Springel} V., 2013, \mnras, 428, 2966

\bibitem[{{Rahmati} {et~al.}(2015){Rahmati}, {Schaye}, {Bower}, {Crain},
  {Furlong}, {Schaller}, \& {Theuns}}]{Rahmati15}
{Rahmati} A., {Schaye} J., {Bower} R.~G., {et~al.}, 2015, ArXiv e-prints,
  astro-ph/1503.05553

\bibitem[{{Rees}(1986)}]{Rees86}
{Rees} M.~J., 1986, \mnras, 218, 25P

\bibitem[{{Rees} \& {Ostriker}(1977)}]{Rees77}
{Rees} M.~J., {Ostriker} J.~P., 1977, \mnras, 179, 541

\bibitem[{{Rela{\~n}o} \& {Kennicutt}(2009)}]{Relano09}
{Rela{\~n}o} M., {Kennicutt} Jr. R.~C., 2009, \apj, 699, 1125

\bibitem[{{Robotham} {et~al.}(2010){Robotham}, {Driver}, {Norberg}, {Baldry},
  {Bamford}, {Hopkins}, {Liske}, {Loveday}, {Peacock}, {Cameron}, {Croom},
  {Doyle}, {Frenk}, {Hill}, {Jones}, {van Kampen}, {Kelvin}, {Kuijken},
  {Nichol}, {Parkinson}, {Popescu}, {Prescott}, {Sharp}, {Sutherland},
  {Thomas}, \& {Tuffs}}]{Robotham10}
{Robotham} A., {Driver} S.~P., {Norberg} P., {et~al.}, 2010, \pasa, 27, 76

\bibitem[{{Rosas-Guevara} {et~al.}(2013){Rosas-Guevara}, {Bower}, {Schaye},
  {Furlong}, {Frenk}, {Booth}, {Crain}, {Dalla Vecchia}, {Schaller}, \&
  {Theuns}}]{RosasGuevara13}
{Rosas-Guevara} Y.~M., {Bower} R.~G., {Schaye} J., {et~al.}, 2013, ArXiv
  e-prints, astro-ph/1312.0598

\bibitem[{{Rosdahl} {et~al.}(2015){Rosdahl}, {Schaye}, {Teyssier}, \&
  {Agertz}}]{Rosdahl15}
{Rosdahl} J., {Schaye} J., {Teyssier} R., {Agertz} O., 2015, ArXiv e-prints,
  astro-ph/1501.04632

\bibitem[{{Sawala} {et~al.}(2014){Sawala}, {Frenk}, {Fattahi}, {Navarro},
  {Theuns}, {Bower}, {Crain}, {Furlong}, {Jenkins}, {Schaller}, \&
  {Schaye}}]{Sawala14}
{Sawala} T., {Frenk} C.~S., {Fattahi} A., {et~al.}, 2014, ArXiv e-prints,
  astro-ph/1406.6362

\bibitem[{{Schaye}(2004)}]{Schaye04}
{Schaye} J., 2004, \apj, 609, 667

\bibitem[{{Schaye} {et~al.}(2015){Schaye}, {Crain}, {Bower}, {Furlong},
  {Schaller}, {Theuns}, {Dalla Vecchia}, {Frenk}, {McCarthy}, {Helly},
  {Jenkins}, {Rosas-Guevara}, {White}, {Baes}, {Booth}, {Camps}, {Navarro},
  {Qu}, {Rahmati}, {Sawala}, {Thomas}, \& {Trayford}}]{Schaye15}
{Schaye} J., {Crain} R.~A., {Bower} R.~G., {et~al.}, 2015, \mnras, 446, 521

\bibitem[{{Schaye} \& {Dalla Vecchia}(2008)}]{Schaye08}
{Schaye} J., {Dalla Vecchia} C., 2008, \mnras, 383, 1210

\bibitem[{{Schaye} {et~al.}(2010){Schaye}, {Dalla Vecchia}, {Booth}, {Wiersma},
  {Theuns}, {Haas}, {Bertone}, {Duffy}, {McCarthy}, \& {van de
  Voort}}]{Schaye10}
{Schaye} J., {Dalla Vecchia} C., {Booth} C.~M., {et~al.}, 2010, \mnras, 402,
  1536

\bibitem[{{Schechter}(1976)}]{Schechter76}
{Schechter} P., 1976, \apj, 203, 297

\bibitem[{{Shen} {et~al.}(2003){Shen}, {Mo}, {White}, {Blanton}, {Kauffmann},
  {Voges}, {Brinkmann}, \& {Csabai}}]{Shen03}
{Shen} S., {Mo} H.~J., {White} S.~D.~M., {et~al.}, 2003, \mnras, 343, 978

\bibitem[{{Sivanandam} {et~al.}(2012){Sivanandam}, {Graham}, {Abraham},
  {Tekatch}, {Steinbring}, {Ngan}, {Welch}, \& {Law}}]{Sivanandam12}
{Sivanandam} S., {Graham} J.~R., {Abraham} R., {et~al.}, 2012, in Society of
  Photo-Optical Instrumentation Engineers (SPIE) Conference Series, Vol. 8446,
  Society of Photo-Optical Instrumentation Engineers (SPIE) Conference Series,
  p.~43

\bibitem[{{Springel}(2005)}]{Springel05}
{Springel} V., 2005, \mnras, 364, 1105

\bibitem[{{Springel} {et~al.}(2001){Springel}, {White}, {Tormen}, \&
  {Kauffmann}}]{Springel01}
{Springel} V., {White} S.~D.~M., {Tormen} G., {Kauffmann} G., 2001, \mnras,
  328, 726

\bibitem[{{Stancliffe} \& {Jeffery}(2007)}]{Stancliffe07}
{Stancliffe} R.~J., {Jeffery} C.~S., 2007, \mnras, 375, 1280

\bibitem[{{Taylor} {et~al.}(2014){Taylor}, {Hopkins}, {Baldry},
  {Bland-Hawthorn}, {Brown}, {Colless}, {Driver}, {Norberg}, {Robotham},
  {Alpaslan}, {Brough}, {Cluver}, {Gunawhardhana}, {Kelvin}, {Liske},
  {Conselice}, {Croom}, {Foster}, {Jarrett}, {Lopez}, \& {Loveday}}]{Taylor14}
{Taylor} E.~N., {Hopkins} A.~M., {Baldry} I.~K., {et~al.}, 2014, ArXiv
  e-prints, astro-ph/1408.5984

\bibitem[{{Thoul} \& {Weinberg}(1995)}]{Thoul95}
{Thoul} A.~A., {Weinberg} D.~H., 1995, \apj, 442, 480

\bibitem[{{Tokunaga} \& {Vacca}(2005)}]{Tokunaga05}
{Tokunaga} A.~T., {Vacca} W.~D., 2005, \pasp, 117, 1459

\bibitem[{{Torrey} {et~al.}(2014){Torrey}, {Snyder}, {Vogelsberger}, {Hayward},
  {Genel}, {Sijacki}, {Springel}, {Hernquist}, {Nelson}, {Kriek}, {Pillepich},
  {Sales}, \& {McBride}}]{Torrey14}
{Torrey} P., {Snyder} G.~F., {Vogelsberger} M., {et~al.}, 2014, ArXiv e-prints,
  astro-ph/1411.3717

\bibitem[{{Tremonti} {et~al.}(2004){Tremonti}, {Heckman}, {Kauffmann},
  {Brinchmann}, {Charlot}, {White}, {Seibert}, {Peng}, {Schlegel}, {Uomoto},
  {Fukugita}, \& {Brinkmann}}]{Tremonti04}
{Tremonti} C.~A., {Heckman} T.~M., {Kauffmann} G., {et~al.}, 2004, \apj, 613,
  898

\bibitem[{{Tully} \& {Fisher}(1977)}]{Tully77}
{Tully} R.~B., {Fisher} J.~R., 1977, \aap, 54, 661

\bibitem[{{Vogelsberger} {et~al.}(2014){Vogelsberger}, {Genel}, {Springel},
  {Torrey}, {Sijacki}, {Xu}, {Snyder}, {Nelson}, \&
  {Hernquist}}]{Vogelsberger14}
{Vogelsberger} M., {Genel} S., {Springel} V., {et~al.}, 2014, ArXiv e-prints,
  astro-ph/1405.2921

\bibitem[{{Walcher} {et~al.}(2011){Walcher}, {Groves}, {Budav{\'a}ri}, \&
  {Dale}}]{Walcher11}
{Walcher} J., {Groves} B., {Budav{\'a}ri} T., {Dale} D., 2011, \apss, 331, 1

\bibitem[{{Wendland}(1995)}]{Wendland95}
{Wendland} H., 1995, Advances Comput. Math., 4, 389

\bibitem[{{White} \& {Frenk}(1991)}]{White91}
{White} S.~D.~M., {Frenk} C.~S., 1991, \apj, 379, 52

\bibitem[{{White} \& {Rees}(1978)}]{White78}
{White} S.~D.~M., {Rees} M.~J., 1978, \mnras, 183, 341

\bibitem[{{Wiersma} {et~al.}(2009{\natexlab{a}}){Wiersma}, {Schaye}, \&
  {Smith}}]{Wiersma09a}
{Wiersma} R.~P.~C., {Schaye} J., {Smith} B.~D., 2009{\natexlab{a}}, \mnras

\bibitem[{{Wiersma} {et~al.}(2009{\natexlab{b}}){Wiersma}, {Schaye}, {Theuns},
  {Dalla Vecchia}, \& {Tornatore}}]{Wiersma09b}
{Wiersma} R.~P.~C., {Schaye} J., {Theuns} T., {Dalla Vecchia} C., {Tornatore}
  L., 2009{\natexlab{b}}, \mnras

\bibitem[{{Worthey}(1994)}]{Worthey94}
{Worthey} G., 1994, \apjs, 95, 107

\bibitem[{{York} {et~al.}(2000){York}, {Adelman}, {Anderson}, {Anderson},
  {Annis}, {Bahcall}, {Bakken}, {Barkhouser}, {Bastian}, {Berman}, {Boroski},
  {Bracker}, {Briegel}, {Briggs}, {Brinkmann}, {Brunner}, {Burles}, {Carey},
  {Carr}, {Castander}, {Chen}, {Colestock}, {Connolly}, {Crocker}, {Csabai},
  {Czarapata}, {Davis}, {Doi}, {Dombeck}, {Eisenstein}, {Ellman}, {Elms},
  {Evans}, {Fan}, {Federwitz}, {Fiscelli}, {Friedman}, {Frieman}, {Fukugita},
  {Gillespie}, {Gunn}, {Gurbani}, {de Haas}, {Haldeman}, {Harris}, {Hayes},
  {Heckman}, {Hennessy}, {Hindsley}, {Holm}, {Holmgren}, {Huang}, {Hull},
  {Husby}, {Ichikawa}, {Ichikawa}, {Ivezi{\'c}}, {Kent}, {Kim}, {Kinney},
  {Klaene}, {Kleinman}, {Kleinman}, {Knapp}, {Korienek}, {Kron}, {Kunszt},
  {Lamb}, {Lee}, {Leger}, {Limmongkol}, {Lindenmeyer}, {Long}, {Loomis},
  {Loveday}, {Lucinio}, {Lupton}, {MacKinnon}, {Mannery}, {Mantsch}, {Margon},
  {McGehee}, {McKay}, {Meiksin}, {Merelli}, {Monet}, {Munn}, {Narayanan},
  {Nash}, {Neilsen}, {Neswold}, {Newberg}, {Nichol}, {Nicinski}, {Nonino},
  {Okada}, {Okamura}, {Ostriker}, {Owen}, {Pauls}, {Peoples}, {Peterson},
  {Petravick}, {Pier}, {Pope}, {Pordes}, {Prosapio}, {Rechenmacher}, {Quinn},
  {Richards}, {Richmond}, {Rivetta}, {Rockosi}, {Ruthmansdorfer}, {Sandford},
  {Schlegel}, {Schneider}, {Sekiguchi}, {Sergey}, {Shimasaku}, {Siegmund},
  {Smee}, {Smith}, {Snedden}, {Stone}, {Stoughton}, {Strauss}, {Stubbs},
  {SubbaRao}, {Szalay}, {Szapudi}, {Szokoly}, {Thakar}, {Tremonti}, {Tucker},
  {Uomoto}, {Vanden Berk}, {Vogeley}, {Waddell}, {Wang}, {Watanabe},
  {Weinberg}, {Yanny}, {Yasuda}, \& {SDSS Collaboration}}]{York00}
{York} D.~G., {Adelman} J., {Anderson} Jr. J.~E., {et~al.}, 2000, \aj, 120,
  1579

\bibitem[{{Zahid} {et~al.}(2014){Zahid}, {Dima}, {Kudritzki}, {Kewley},
  {Geller}, {Hwang}, {Silverman}, \& {Kashino}}]{Zahid14}
{Zahid} H.~J., {Dima} G.~I., {Kudritzki} R.-P., {et~al.}, 2014, \apj, 791, 130

\bibitem[{{Zaragoza-Cardiel} {et~al.}(2014){Zaragoza-Cardiel}, {Font},
  {Beckman}, {Garc{\'{\i}}a-Lorenzo}, {Erroz-Ferrer}, \&
  {Guti{\'e}rrez}}]{Zaragoza14}
{Zaragoza-Cardiel} J., {Font} J., {Beckman} J.~E., {et~al.}, 2014, \mnras, 445,
  1412

\end{thebibliography}

\appendix

\section{Re-sampling}

\label{sec:resmp}
\label{ap:adapt}

\begin{figure*}
  \centering
  \includegraphics[width=0.49\textwidth]{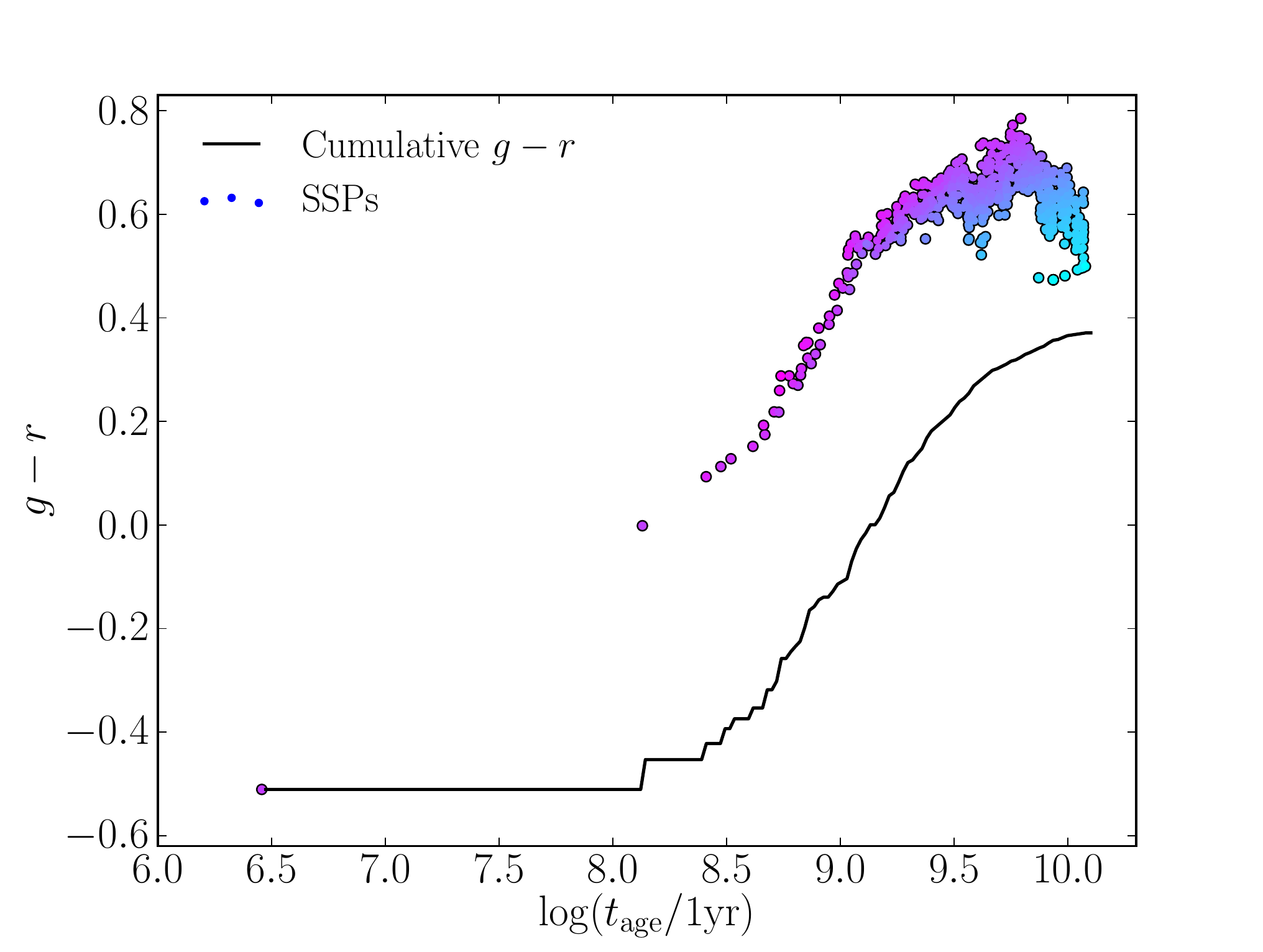}
  \includegraphics[width=0.49\textwidth]{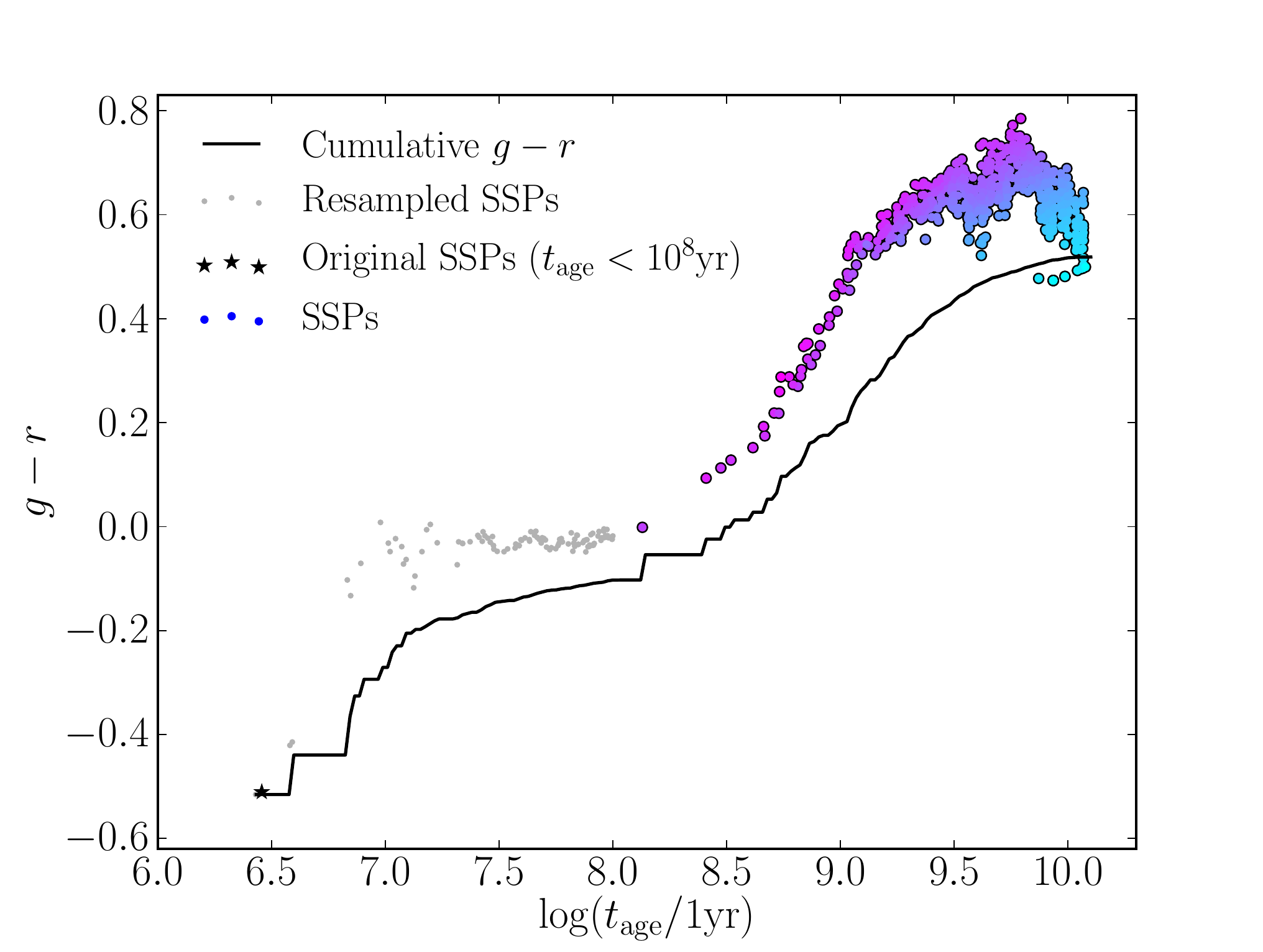}
  \caption{Demonstration of the re-sampling technique for the
    stellar component of an example \eagle{} galaxy from the 50$^3$
    Mpc$^3$ box at redshift $z = 0.1$. The left panel shows particle SSP
    $g-r$ colours as a function of starburst age for an individual
    \eagle{} galaxy as points coloured from blue to magenta indicating
    low to high metallicity. The black line indicates the cumulative
    galaxy colour when including particles with increasing age (from
    left to right) so that the far right point of the line gives the
    total galaxy colour. The right panel shows the same galaxy after
    star particles of age $< 100$ Myr have been re-sampled. These particles
    (black stars; only one in this example) are removed from the photometry calculation and the
    re-sampled stars (grey points) are included. This changes the
    cumulative $g-r$ indicated by the black line, and yields a
    different overall colour.}
    \label{fig:resamp}
\end{figure*}

Young stellar populations are optically much brighter and
bluer than older populations,
due to the presence of O, B and A type stars. As such, the convergence of optical magnitudes and
colours depend on wavelength and star formation in a complex
way. Coarse sampling of young star particles is expected to introduce
Poisson scatter in the colour distribution of low-mass galaxies. 
The standard resolution particle mass of $m_g = 2 \times 10^6 {\rm
  {\rm M}_\odot}$ is $\sim$2 orders of magnitude larger than individual H{\sc ii}
regions \citep{Zaragoza14, Relano09}, so this scatter may be
artificially high in the simulations compared to observed galaxies.

To mitigate this effect, we re-sample recent star formation in
the simulation outputs at finer mass resolution than the simulation gas particle
mass. The re-sampling procedure assumes that the galactic star
formation rate is constant over the previous 100 Myr of galaxy
evolution. The general procedure is outlined below.

Gas particles with non-zero star formation rates and star
particles that formed within the past 100 Myr are first identified.
Star formation rates of progenitor gas particles are then obtained
for the selected star particles. We
calculate the star formation rates for star particles using the
stored gas density at birth and initial particle masses \citep[see][]{Schaye04}, and assume that the progenitor
particles lie on the equation of state (see S15). We then take the sum of the star
formation rates found for the gas and young stars to be the galactic
star formation rate.

We then use some discretisation mass, $m_{\rm dis}$, to represent
the mass of re-sampled star particles. Each selected star or gas particle
is split into the integer number of equal-mass subparticles
yielding a subparticle mass closest to $m_{\rm dis}$. 
We also obtain the ratio of subparticle mass over progenitor particle star formation
rate and interpret this as a conversion timescale for each
subparticle to become a star particle. We then randomly sample individual conversion
times $t$ for subparticles using an exponential distribution of the
appropriate timescale. If $t < 100$ Myr, a subparticle is deemed to
be converted into a star particle with $t_{\rm age} = t_i$, a mass
equal to the subparticle mass and the $Z_\star$ value inherited from
the parent particle.

For our analysis we use $m_{\rm dis} =  10^4$ M$_\odot$ as our
target re-sampling mass resolution. This represents a more reasonable H{\sc
  ii} region mass, to better reproduce the discretisation in observed
galaxies. To illustrate the effect of re-sampling,
the left-hand panel of Figure \ref{fig:resamp} shows the $g-r$ colour and
age of individual star particles in an actively star-forming \eagle{} galaxy. We also
plot the cumulative colour of the galaxy by star particle age as a
black line. This shows the influence that the few youngest star
particles can have on the colour of the entire galaxy. The left and right
panels show the result  with and without re-sampling respectively, yielding
different overall colours for the galaxy (seen as the rightmost point
of the black line). The presence of a single very young star particle
in the simulation output causes this galaxy to appear $\sim 0.2$ mag bluer
than with re-sampling. Though re-sampling can have a significant effect
on individual galaxy colours, the colour distributions for the entire
\eagle{} population are only marginally affected. At the low-mass end, the
re-sampling generally serves to tighten the $g-r$ colour distribution,
move the blue peak to slightly bluer colours ($\sim0.05$ mag) and
to suppress extremely blue outliers.

\section{Colour Convergence}
\label{sec:conv}

\begin{figure}
  \centering
  \includegraphics[width=0.98\columnwidth]{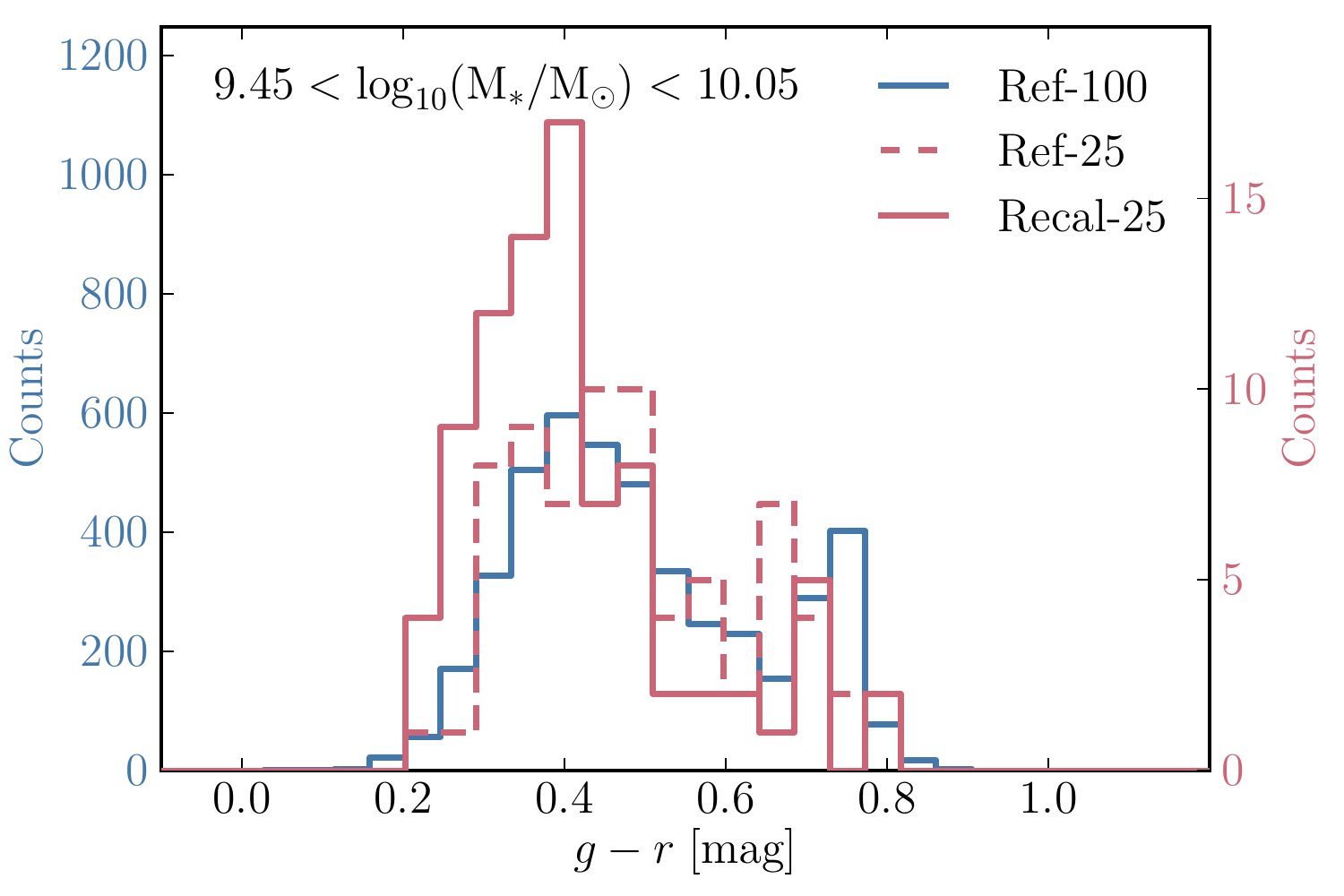}
  \includegraphics[width=0.98\columnwidth]{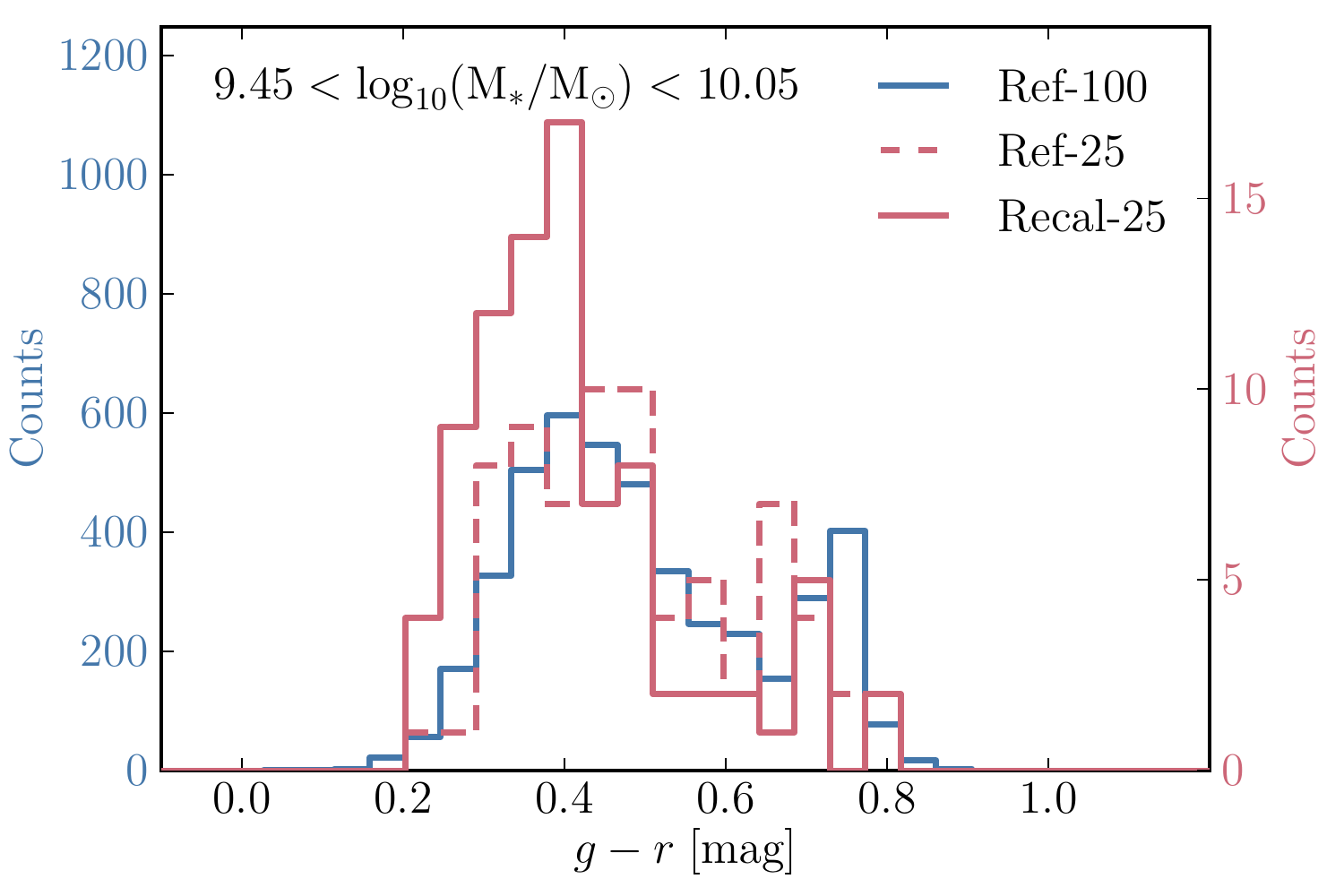}
  \caption{Comparison of $g-r$ colour distributions of the \ac{Ref-100}, \ac{Ref-25} and
   \ac{Recal-25} simulations (see table \ref{tab:sims}) to assess the weak
    convergence and volume effects on model colours. \ac{Ref-100} and
    \ac{Ref-25} have the same resolution, while \ac{Recal-25} has a
    resolution 8 times finer. The top and bottom
    panels show galaxy stellar mass ranges of $9.45 < \log({\rm M}_\ast/{\rm
      M}_\odot) \leq 10.05$ and $8.7 < \log({\rm M}_\ast/{\rm
      M}_\odot) \leq 9.3$, respectively. In both panels, the blue and red
    histograms represent the counts per colour bin in the 100 Mpc and
  25 Mpc simulations respectively. The \ac{Recal-25} and \ac{Ref-25}
  simulations, are plotted as solid and dashed lines,
  respectively. Separate $y$-axes are labelled and coloured to
  correspond to the two volumes, with their ranges scaled by a
  factor of 64 to account for the differing box volume. Both
  resolution and box size appear to significantly effect the colour
  distributions of low-mass \eagle{} galaxies.}
\label{fig:conv}  
\end{figure}

As simulation \ac{Recal-25} has a factor of 8 finer mass resolution
than the fiducial model \ac{Ref-100}, the stellar mass threshold above which
galaxies are considered well resolved is pushed to lower
masses. By comparing colour distributions of the \ac{Ref-100},
\ac{Ref-25} and \ac{Recal-25} simulations (Table~\ref{tab:sims}, S15) for
galaxies within a certain mass range, we attempt to decouple the
effects of simulation volume and resolution on the colours of low-mass
galaxies in \eagle{}. In Figure
\ref{fig:conv} we compare colour distributions for galaxies of mass
$9.45 < \log({\rm M}_\ast/{\rm M}_\odot) \leq 10.05 $ and  $8.7 <
\log({\rm M}_\ast/{\rm M}_\odot) \leq 9.3$ in the top and bottom
panels respectively. The histograms for differing simulation volumes have
different y axis ranges, with the 25 Mpc simulation axis range a factor of 64
smaller to account for the differing simulation volumes.

In the $9.45 < \log({\rm M}_\ast/{\rm M}_\odot) \leq 10.05$ mass
range, the position of the red and blue peaks appear roughly the same
in the different simulations. However, the relative strengths of the red and blue
populations differ, with the red sequence being
significantly weaker than the blue cloud in the high-resolution \ac{Recal-25}
model compared to \ac{Ref-100} and \ac{Recal-25}. This is consistent with the
lower passive fractions in the high resolution simulation at $z=0.1$
shown in S15. 

The $8.7 < \log({\rm M}_\ast/{\rm  M}_\odot) \leq 9.3$ range shows
less consistency, with the red  sequence becoming practically absent
in the \ac{Recal-25} model while 
remaining in the \ac{Ref-100} and \ac{Ref-25} models. The redder colour
and larger scatter of the blue population in the reference model is
attributable to poor sampling of star forming gas in these galaxies. The
lower star formation rates in the fiducial volume may also account for
the different colours. However, we also see a larger difference between
\ac{Ref-25} and \ac{Ref-100} here, particularly in the
relative contributions of the red and blue populations. We
attribute the higher contribution of the red sequence in the \ac{Ref-100} model to the presence of large cluster environments in the
\ac{Ref-100} simulations, and thus quenched satellite galaxies, that
are not sampled by the \ac{Ref-25} box. This suggests that volume
effects also contribute to the weaker red sequence seen in the
\ac{Recal-25} box. In both plots the greater area under
the \ac{Recal-25} histogram is indicative of the systematic shift in
galaxy number densities between the simulations, also seen in Figure
\ref{fig:lfs}.

\section{SSP Parameter Influence}
\label{ap:gallazzi}

As intrinsic galaxy colours are sensitive to star formation histories
and elemental abundance patterns, comparing \eagle{} model colours
directly to observed galaxy colours is a difficult way to disentangle
the influence of different SSP parameters and to identify the source of
any discrepancies.

To go some way towards assessing how the \eagle{} stellar metallicities and star
formation histories influence our mock photometry, we use the simple
photometric model without dust (\modela{}).
Two sets of photometric data are first generated for the simulated
galaxy sample using simulation output for one parameter while using
empirical relations for the other. The galaxy metallicities and
light-weighted ages (LWAs) as functions of stellar mass presented by
\citet{Gallazzi05} are used to provide the empirical
input. For the LWA values, we include a Gaussian scatter about the
median values of the published width, which is assumed to be
uncorrelated with metallicities. Clearly the
assumption that galaxies may be treated as a single starburst and that
the metallicity and age parameters are uncorrelated are poor, so the
amount of information that can be drawn from this type analysis is
limited. These plots serve as a basic qualitative illustration of the
influence of different SSP parameters on galaxy colours.

\begin{figure}
  \centering
  \includegraphics[width=0.98\columnwidth]{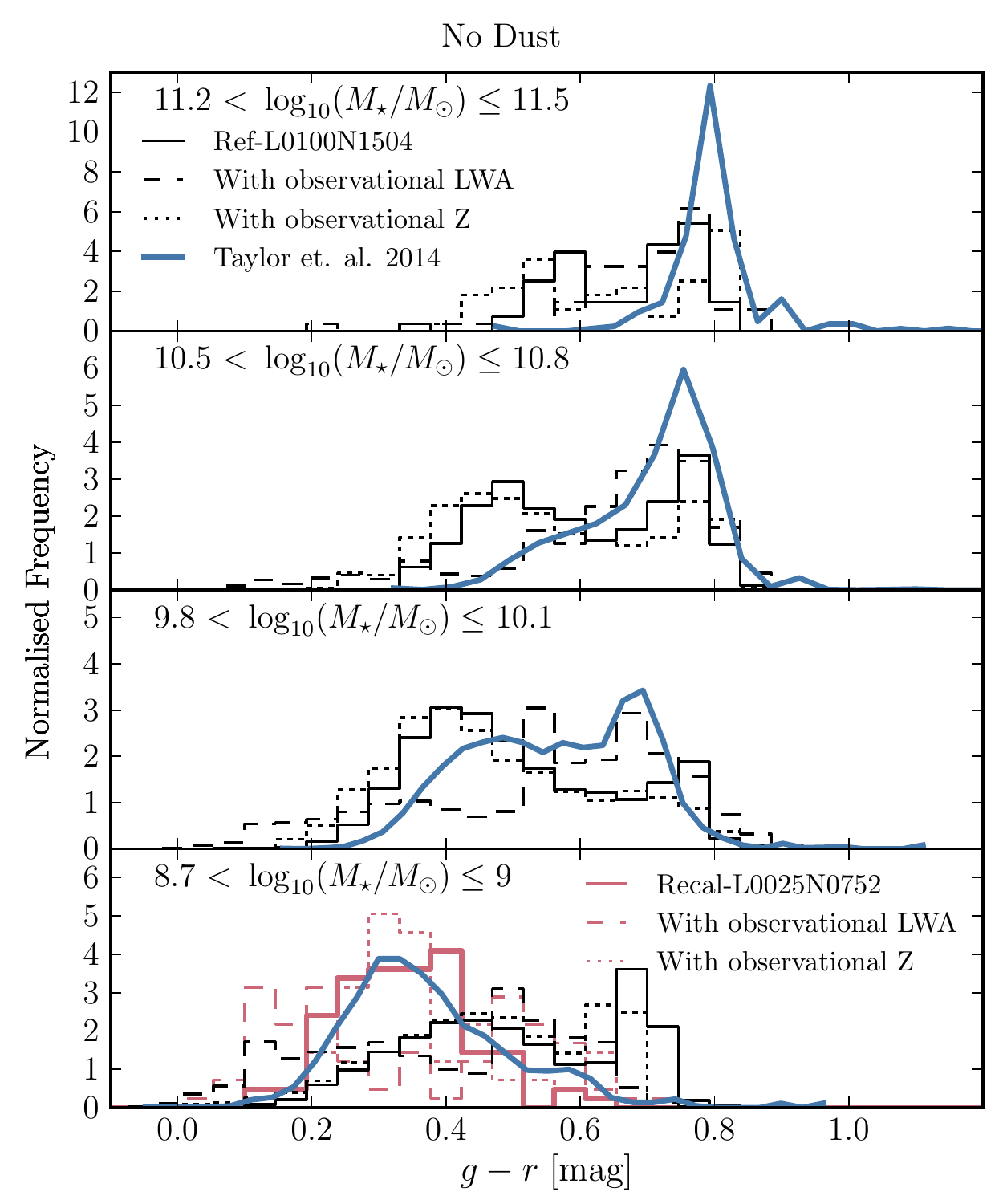}
    \caption{$g-r$ colour distributions for \eagle{} galaxies for the
      dust-free model. The 4 panels show
      colour distributions for 4 bins of stellar mass, as
      indicated by the legend. The solid, dotted and dashed lines show
      the \eagle{} SSP values and the \eagle{} SSP values with Gallazzi et al.
      (2005) metallicities and ages respectively. Gallazzi
      metallicities and ages are
      assigned to each galaxy, based on the median parameter values at
      the galactic stellar mass. $Z$ values are simply taken as the
      observed median value. The LWA values are sampled from a normal
      distribution with standard deviation taken from Gallazzi et. al.
      (2005), assuming that the scatter in age and metallicity is
      uncorrelated. We see that the complex star formation histories
      of \eagle{} provide a better match to the observed colour
      distributions than a single SSP model using empirical values for
    age and metallicity.} 
    \label{fig:gallazzi}
\end{figure}

The colour distributions in four $M_\ast$ bins are plotted for
the simulation-empirical hybrid photometry models, and are compared to the
\eagle{} photometry in Figure \ref{fig:gallazzi}. The black
lines indicate distributions of \ac{Ref-100} simulation galaxies.
The distributions using the raw emission model with \eagle{} ages and
metallicities are plotted as solid histograms. The photometry models using
observed LWA and $Z_\ast$ values are plotted as the dashed and
dotted lines, respectively. The observational data of \citet{Taylor14}
is also plotted in blue for comparison.

We see that the age parameter has the biggest influence on the colour
distribution, with the empirical ages introducing a generally larger spread
than metallicities, when compared to the pure \eagle{}
photometry. We have verified that this is still the case when we
include the scatter on observed metallicity values. Figure \ref{fig:gallazzi} shows that giving galaxies
a single age stellar population using the observational LWA data of
\citet{Gallazzi05} (dashed line) works reasonably well in the two most
massive bins where stellar populations are old. However in the
lower-mass bins where galaxies are generally younger they provide a
poor fit to the observed colours, inferior to our  model photometry using
the complex star formation histories of \eagle{} (solid line).

The bimodality seen for the full \eagle{} photometry in the two most
massive bins, but not for in the observational LWA model, shows that the
\eagle{} populations are intrinsically bimodal in age. This supports the
assertion that there is an excess of star forming galaxies in this
regime relative to the observed population. The bluer than observed
high-$M_\star$ red sequence in the observational LWA model could be a
result of the lower metallicities of high-$M_\star$ galaxies. The
inferior agreement of the observational LWA model relative to the full
\eagle{} model in the lower-mass bins suggests that the complex star
formation histories of \eagle{} reproduce the data better than an
empirical model assuming a single age population. 

The observational $Z$ model reveals a poor fit to observation for the
two highest-mass bins. The red sequence is also much less prominent
than seen in the observations and the other models across the $M_\ast$
range.

The systematic effect of assuming uncorrelated scatter between the age parameter may also account for the fact that the
colour distributions are broader and flatter than observed for this
model, especially in the low-mass bins.   

The resolution effects that drive much improved agreement
between observed low-mass colours and \ac{Recal-25} relative to \ac{Ref-100}  
are noted in section~\ref{sec:results} and appendix~\ref{sec:conv}.
In the lowest-mass bin of Figure \ref{fig:gallazzi}, we see that
using observed metallicities has less impact on \eagle{} colours than
using observed LWAs. This indicates that star formation rate
resolution is the primary resolution effect on colours, with
metallicity resolution secondary to this. The presence of a faint red
sequence is due to lower star formation rates and higher stellar ages
than found in low-mass \ac{Ref-100} galaxies, whereas the position of the red
sequence is redder by $\sim 0.1$ due to the higher than observed
metallicities at these masses. The star formation rate resolution
is also the main contributor to the redder than observed blue cloud
position in \ac{Ref-100}.

\end{document}